\documentclass[sigconf, natbib=false]{acmart}
\AtBeginDocument{%
  }

\copyrightyear{2025}
\acmYear{2025}
\setcopyright{licensedothergov}\acmConference[KDD '25]{Proceedings of the 31st ACM SIGKDD Conference on Knowledge Discovery and Data Mining V.1}{August 3--7, 2025}{Toronto, ON, Canada}
\acmBooktitle{Proceedings of the 31st ACM SIGKDD Conference on Knowledge Discovery and Data Mining V.1 (KDD '25), August 3--7, 2025, Toronto, ON, Canada}
\acmDOI{10.1145/3690624.3709405}
\acmISBN{979-8-4007-1245-6/25/08}

\RequirePackage[
  datamodel=acmdatamodel,
  style=acmnumeric,
    natbib=false, 
]{biblatex}

\usepackage{stackengine} 
\usepackage{cleveref} 
\usepackage{multirow} 
\usepackage{rotating} 
\usepackage{booktabs} 
\usepackage{arydshln} 
\usepackage{svg} 
\usepackage{subcaption} 
\usepackage{array} 

\newcommand{\x}{\mathbf{x}} 
\def\myeq{\mathrel{\ensurestackMath{\stackon[1pt]{=}{\scriptscriptstyle\Delta}}}} 

\begin{document}

\title{ECGrecover: A Deep Learning Approach for Electrocardiogram Signal Completion}


\author{Alex Lence}
\authornotemark[1]
\email{alex.lence@ird.fr}
\author{Federica Granese}\authornotemark[1]
\author{Ahmad Fall}
\authornote{Equal contributions.}

\affiliation{
  \institution{IRD, Sorbonne Université, Unité de Modélisation Mathématique et Informatique des Systèmes Complexes,  UMMISCO, F-93143, Bondy, France}
  \city{Bondy}
  \country{France}
  }

\author{Blaise Hanczar}
\affiliation{%
  \institution{IBISC, University Paris-Saclay, University Evry}
  \city{Evry}
  \country{France}
  }

\author{Joe-Elie Salem}
\affiliation{
  \institution{Clinical Investigation Center Paris-Est, CIC-1901, INSERM, UNICO-GRECO Cardio-Oncology Program, Pitié-Salpêtrière University Hospital, Sorbonne Université}
  \city{Paris}
  \country{France}
  \institution{Vanderbilt University Medical Center}
  \city{Nashville}
  \country{USE}
  }

\author{Jean-Daniel Zucker}
\author{Edi Prifti}
\authornotemark[2]
\email{edi.prifti@ird.fr}
\affiliation{
  \institution{IRD, Sorbonne Université, Unité de Modélisation Mathématique et Informatique des Systèmes Complexes,  UMMISCO, F-93143, Bondy, France}
  \city{Bondy}
  \country{France}
  \institution{INSERM, Nutrition et Obesities; systemic approaches, NutriOmique, AP-HP, Hôpital Pitié-Salpêtrière}
  \city{Paris}
  \country{France}
  }
\authornote{Corresponding author.}
\renewcommand{\shortauthors}{Alex Lence et al.}

\begin{abstract}
In this work, we address the challenge of reconstructing the complete 12-lead ECG signal from its incomplete parts. We focus on two main scenarios: (i) reconstructing missing signal segments within an ECG lead and (ii) recovering entire leads from signal in another unique lead. Two emerging clinical applications emphasize the relevance of our work. The first is the increasing need to digitize paper-stored ECGs for utilization in AI-based applications, often limited to digital 12 lead 10s ECGs. The second is the widespread use of wearable devices that record ECGs but typically capture only one or a few leads. In both cases, a non-negligible amount of information is lost or not recorded. Our approach aims to recover this missing signal. We propose ECGrecover, a U-Net neural network model trained on a novel composite objective function to address the reconstruction problem. This function incorporates both spatial and temporal features of the ECG by combining the distance in amplitude and sycnhronization through time between the reconstructed and the real digital signals. We used real-life ECG datasets and through comprehensive assessments compared ECGrecover with three state-of-the-art methods based on generative adversarial networks (EKGAN, Pix2Pix) as well as the CopyPaste strategy. The results demonstrated that ECGrecover consistently outperformed state-of-the-art methods in standard distortion metrics as well as in preserving critical ECG characteristics, particularly the P, QRS, and T wave coordinates.

\end{abstract}





\maketitle

\section{Introduction}
\label{sec:introduction}

An electrocardiogram (ECG) is a diagnostic test broadly used to evaluate cardiac health. It is commonly performed by placing electrodes at specific locations on the patient's body to capture the heart's electrical signals. The \textit{thoracic} electrodes (V1-V6) provide information on the right and left ventricles, while the \textit{limb} electrodes (I, II, III, aVR, aVL, aVF) offer insights into the heart's electrical signals in various spatial directions~\cite{parmet2003electrocardiograms}. Healthcare professionals can effectively diagnose and assess cardiac conditions by leveraging the combined information obtained from these electrodes. However, depending on the specific lead under examination, certain diseases may manifest with varied expressions on an ECG. For instance, when \textit{myocardial infarction} occurs in the inferior part of the heart, it produces alterations on the ECG signal that are more visible in the leads II, aVF, and III~\footnote{\url{https://ecgwaves.com/topic/localization-localize-myocardial-infarction-ischemia-coronary-artery-occlusion-culprit-stemi/}}.
\begin{figure}
    \centering
    \includegraphics[width=0.8\columnwidth]{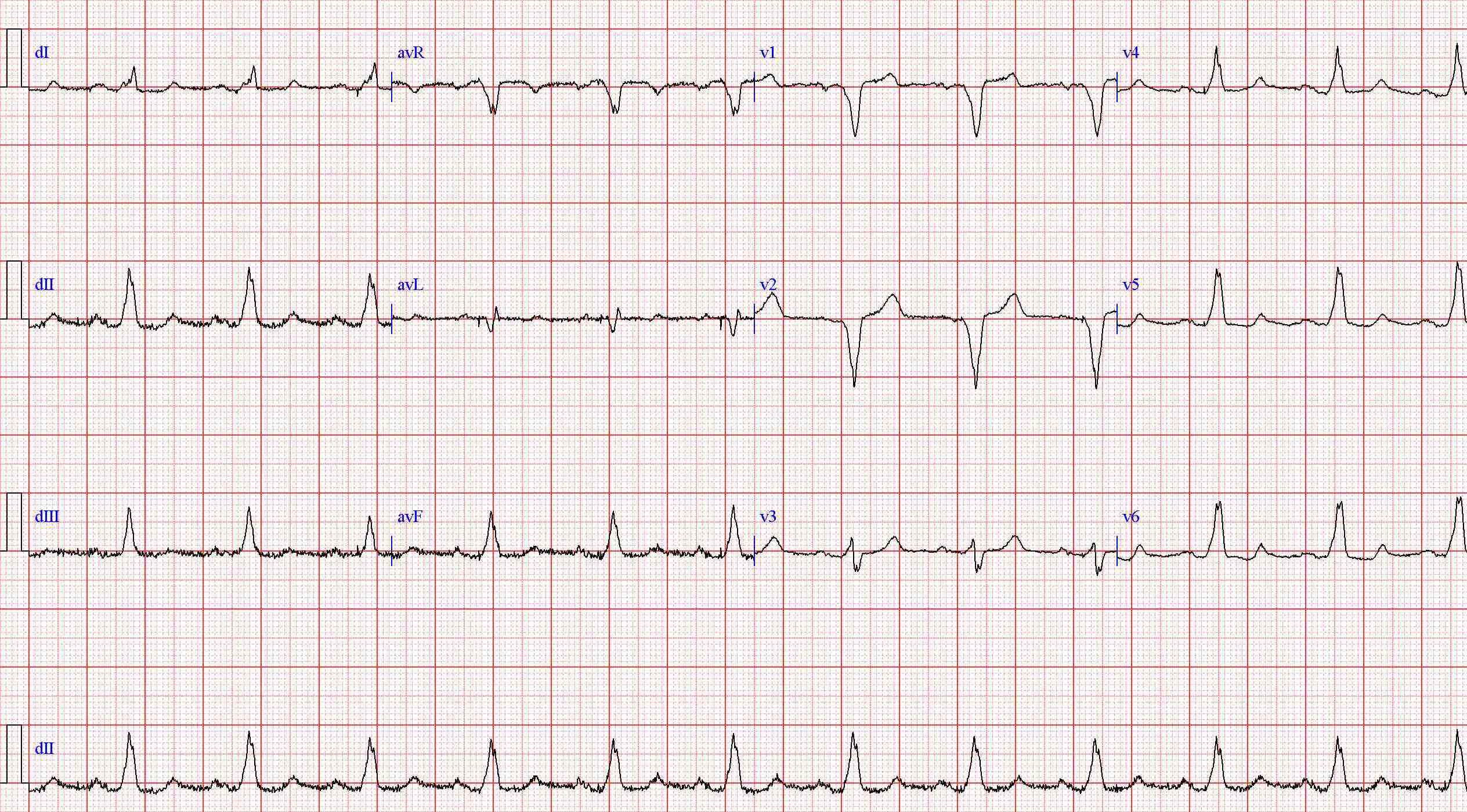}
    \caption{Example of a 12-lead ECG. The machine has recorded a 10-second complete recording for all 12 leads, but in the PDF and paper printed format, only 2.5 seconds are kept for all the leads, in addition to the complete copy of lead II. This is illustrated by the configuration C3 in~\cref{fig:configurations}).}
    \label{fig:ecg_example}
\end{figure}

While a comprehensive disease diagnosis requires all the leads (in most cases, 12), novel technological advancements have revolutionized how individuals monitor their health. ECGs are frequently recorded with wearable devices for longer periods (in contrast with the classical clinical exam, which lasts 10-30 seconds). However, such devices measure only one or a few leads in the best cases, e.g., certain smartwatches can record and transmit a single-lead ECG, equivalent to lead I~\cite{van2023using}. 

Moreover, despite the ongoing shift towards AI and automated ECG analyses, the majority of healthcare establishments around the world, particularly in developing countries, still rely on paper to record and store ECG~\cite{walker2009low}. This poses obvious accessibility issues and makes computer-based analyses especially challenging. In the most common formats, each lead is only partially represented (i.e., only 2.5 or 5 seconds). Although several tools exist~\cite{lence_automatic_2023} to digitize the ECG stored on paper, they cannot compensate for the missing signal. Relying on a partial signal for a medical assessment can lead to even more catastrophic consequences. For example, QT alteration involves fluctuations in the duration of the QT interval. Analyzing a longer segment may provide a clearer picture of QT alteration and other subtle changes that might not be visible in shorter segments. Conditions such as bradycardia or extrasystoles require practitioners to analyze more than three heartbeats to determine the presence of the pathology. The significance of QT alterations lies in its association with an increased risk of ventricular arrhythmias, including Torsades-de-Pointes (TdP), which can degenerate into potentially life-threatening ventricular fibrillation~\cite{crotti2008congenital, prifti2021deep, salem2017genome}

Consequently, reconstructing missing portions of an ECG signal, whether it involves an entire lead or part of it, is a challenging task with implications for both the medical field and computer science. To approach this problem, we formally define it and address it from two perspectives: \textit{ECG lead-reconstruction} and \textit{ECG segment recovery}. In ECG lead reconstruction, the purpose is to reconstruct the 11 missing leads using the full trace of only one lead, while for ECG segment recovery, the goal is to recover the entire 10-second signal for all the leads from only partial existing segments.

In~\Cref{fig:ecg_example}, we show a 12-lead ECG recorded in paper format. The recordings from the various leads are concatenated according to the standard 3x4 printing format (3 partial channels with 4 partial leads each, starting with lead I, followed by leads aVR, V1, and V4, and a fourth channel containing a complete lead, either lead II or I). Lastly, the reference pulses are used for amplitude calibration at the beginning of each row. We proposed a masking approach involving specific masks to replicate real-life scenarios like those illustrated in~\Cref{fig:ecg_example}. Given a 12-lead ECG input, the masks are functions that return the same signal, with parts of it replaced by random values drawn from a uniform distribution. For instance, the 3x4 format corresponds to the C3 mask  in~\Cref{fig:configurations}, where the portion of the signal retained is in green, which we refer to as the \textit{primer}.

Our approach involved developing a U-Net-like neural network model, trained with a novel original objective function, combining the mean squared error loss with the Pearson correlation loss. The former serves to guide the model in preserving the magnitude of the original ECG signal, while the latter allows the model to maintain synchronization with the original signal. We named this approach \textit{ECGrecover}. The final model was trained on a dataset comprising 4498 real-life 12-lead, 10-second ECGs from a cohort of 990 healthy volunteers ~\cite{salem2017genome} who were administered 80mg of Sotalol, a QT prolonging drug. During training, all the masks depicted in~\Cref{fig:ecg_example} were applied to each ECG of the dataset. However, each mask was applied separately to the testing set during the evaluation.

The main contributions of this work are threefold:\\
\noindent (1) \textbf{
We identified 17 realistic \textit{scenarios} where ECG signals may be partially available}. Our analysis differentiates between two cases: reconstructing missing segments within an ECG lead and reconstructing the entire set of 12 leads from one single lead. In addition, a formal definition of the problem is provided (\Cref{sec:problem_formulation}).\\
\noindent (2) We introduce ECGrecover\footnote{Code available at \url{https://github.com/UMMISCO/ecgrecover}}, a U-Net model for segment-recovery and lead-reconstruction tasks. \textbf{ECGrecover is trained with a novel objective function combining the mean squared error and the Pearson correlation} to ensure the reconstructed signal matches the original in amplitude and preserves synchronized waveform patterns, crucial for clinical applications (\Cref{sec:ecg_recover}).\\
\noindent (3) \textbf{In close collaboration with expert cardiologists, we extensively evaluated ECGrecover on a dataset composed of real-life ECG}~\cite{salem2017genome,prifti2021deep}. Besides considering classical distortion metrics, this assessment also focused on the model's ability to preserve clinically relevant signal components of the ECG signal, such as QT distance (\Cref{sec:experiments}). 
\raggedbottom

\subsection{Related works}
\label{sec:related}
Synthesizing ECG signals involves creating artificial waveforms that resemble real ECG patterns. Generally, ECG synthesis has been studied through the lens of data augmentation~\cite{rahman2023systematic} aiming to increase the size of the training dataset by generating "authentic"-like ECGs conditioned or not on heart pathologies, to enhance classifier performance in arrhythmia predictions (e.g., \cite{chen2022me, golany2020improving, golany2019pgans}); to improve ECG quality assessment~\cite{zhou2021electrocardiogram}, and to improve ECG anomaly detection~\cite{li2022contrastive}, among other applications. Existing techniques often involve training deep neural networks such as Variational Autoencoders (VAE) (e.g.~\cite{soleimani2022enhancing, kuznetsov2021interpretable}), Generative Adversarial Networks (GAN) (e.g.~\cite{ye2019ecg,hazra2020synsiggan}), or CycleGAN (e.g~\cite{abdelmadjid2022neural}).

ECG completion has been addressed in~\cite{atoui2010novel, wang2019novel}, where the remaining leads are reconstructed from three other complete leads (I, II, and V2). In~\cite{cho2020artificial, lee2015reconstruction}, the limb leads (frontal electrodes) are used to generate the thoracic leads (precordial electrodes). Nonetheless, these studies do not address the scenario where the reconstruction of the remaining leads is based only on a single lead, nor when the leads are partially recorded (less than 10 seconds).

Another solution is proposed in~\cite{joo2023twelve}, where the authors introduced EKGAN, a conditional GAN based on Pix2Pix~\cite{isola2017image}, consisting of two generators and one 1D U-Net discriminator~\cite{ronneberger2015u}, that tackles the reconstruction of 12-lead ECGs from one single lead (lead I). Prior attempts in synthesizing multiple ECGs from a single-lead ECG were made in~\cite{seo2022multiple} where the authors proposed a GAN composed of a U-Net generator and a discriminator based on the PatchGAN~\cite{isola2017image} architecture. The model is designed to take 2.5-second ECG segments from lead I as input and reconstruct 2.5-second segments for any other leads. The authors trained 11 different models, each tailored for reconstructing the ECG signal of a specific lead from the input derived from lead I.

It is worth noting that in both~\cite{joo2023twelve} and~\cite{seo2022multiple}, the experimental data comprise 10s ECG recordings at 500Hz. However, the evaluation is conducted on different databases: a private one for~\cite{joo2023twelve} and PTBXL~\cite{wagner2020ptb} along with a second dataset~\cite{zheng202012} for~\cite{rautaharju1981exploitation, seo2022multiple}.

Finally, a more naive technique for the ECG segment-recovery problem involves copy-pasting the available signal to fill up the missing segments, as suggested by Badilini et al. in~\cite{badilini2005ecgscan}. 

Considering a broader context beyond just ECGs, Alcaraz et al.~\cite{alcaraz2022diffusion} propose a time series imputation model that integrates the principles of classical conditional time diffusion models with the methodologies of structured state space models. Unlike our setting, which focuses on 17 realistic cases, they focus on scenarios not necessarily connected to our analysis. Specifically, they address random missing, where zero entries of an imputation mask are sampled randomly across all channels; random block missing (RBM), where one segment per channel is sampled as the imputation target; blackout missing (BM), where a single segment is missing across all channels; and time series forecasting (TF), a special case of BM where the imputation region spans $t$ time steps at the end of the sequence.  
While Alcaraz et al. provide experiments on the PTB-XL datasets, their method cannot be fairly compared to ours due to key differences. Their ECGs are downsampled from 500 Hz to 100 Hz, which could shift data distributions significantly. Additionally, their model's training time is considerably longer at 36 hours, affecting its practicality and scalability. 
Finally, we would like to point out that, in our configurations, the models must reconstruct up to 92\% of the original signal in the worst-case scenario (C1), whereas the proposed schemes in~\cite{alcaraz2022diffusion} only require up to 50\% of the signal to be generated each time (RBM setting).

\section{Methods}
In this section, we formally define the problem of ECG reconstruction. Subsequently, we introduce ECGrecover, our proposed approach, and extensively evaluate it while comparing it with state-of-the-art methods using relevant real-life ECG datasets. In \Cref{tab:symbols}, the list of the symbols used throughout this article.
\label{sec:preliminaries}
\subsection{Problem formulation}
\label{sec:problem_formulation}
Let $X\sim P_{X}$ be the random variable (r.v.) for which we have realization $\x\in\mathcal{X}\subseteq{R}^{L\times N}$, where $L\in\mathbb{N}$ represents the number of leads and $N\in\mathbb{N}$ the length of the signal in terms of sample points. We will refer to $\mathcal{X}$ as the set of \textit{valid} ECGs, meaning they make sense in clinical terms. In this specific context, we say that an ECG is valid if it is also \textit{complete} i.e., for all $l\in\{1, \dots, L\}$ and $n\in\{1, \dots, N\}$, $\x_{ln}\neq0$, where $\x_{ln}$ indicates the specific data point at lead $l$ and time point $n$ within the ECG matrix, while $\x_{l}$ denotes the ECG vector of $N$ time points at lead $l$. We call it \textit{incomplete} otherwise, and we denote the set of incomplete ECGs as $\widetilde{\mathcal{X}}\subset\mathbb{R}^{L\times N}\setminus\mathcal{X}$.

\begin{table}[t]
    \caption{Table of symbols}
    \centering
    \begin{tabular}{rl}
    \toprule
         Symbol & Definition \\
    \midrule
    $L$ & Number of leads in ECG\\
    $N$ & Length of the ECG\\ 
    $\x\in\mathcal{X}\subset\mathbb{R}^{L\times N}$ & Valid (and complete) multi-lead ECG\\ 
    $\widetilde{\x}\in\widetilde{\mathcal{X}}\subset\mathbb{R}^{L\times N}\setminus\mathcal{X}$ & Incomplete multi-lead ECG\\
    $f_\theta(\cdot)$ & Model parameterized by $\theta$\\
    $\widehat{\x}\equiv f_\theta(\widetilde{\x})$ & Reconstructed multi-lead ECG\\
    $\x_{l}$, $\widetilde{\x}_{l}$, $\widehat{\x}_{l}\in\mathbb{R}^{1\times N}$ & ECG signal at lead $l$\\
     $\x_{ln}$, $\widetilde{\x}_{ln}$, $\widehat{\x}_{ln}\in\mathbb{R}$  & ECG value at lead $l$ and point $n$\\
     $\mathcal{L}_{\text{MSE}}(\cdot)$ & Mean Squared Error loss\\
     $\mathcal{L}_{\text{Pearson}}(\cdot)$ & Pearson loss\\
      & Sample correlation coefficient\\ $r_{lj}$ & (i.e., Pearson correlation coefficient)\\&between lead $l$ and $j$\\
    \bottomrule
    \end{tabular}
    \label{tab:symbols}
\end{table}

\begin{figure*}[t]
	\centering
		\begin{subfigure}[b]{\columnwidth}
		\centering
		\includegraphics[width=.85\columnwidth]{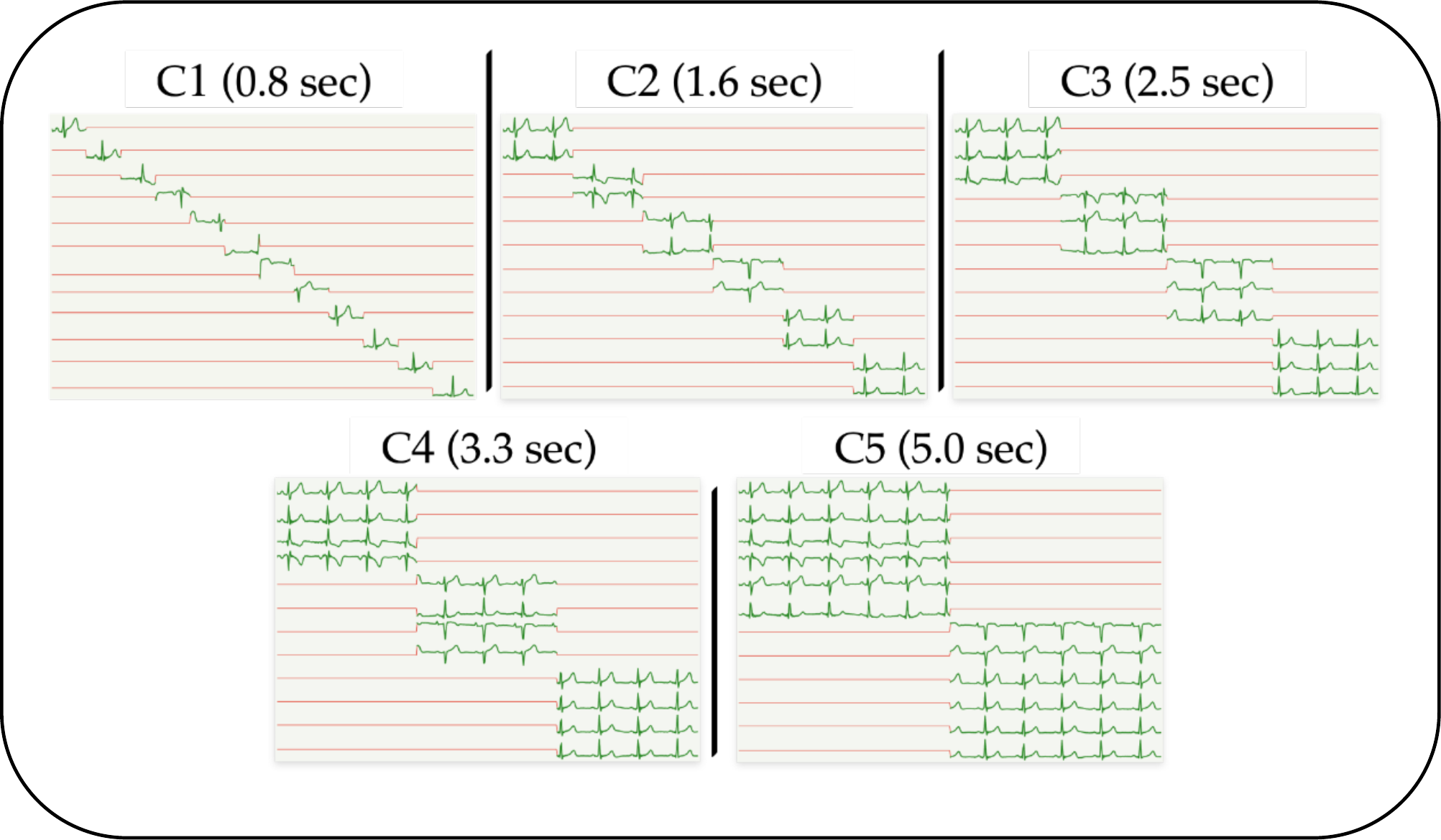}
		\caption{Masks to simulate ECG segment-recovery.}
		\label{fig:segment_mask}
	\end{subfigure}
 \hfill
		\begin{subfigure}[b]{\columnwidth}
		\centering
		\includegraphics[width=.85\columnwidth]{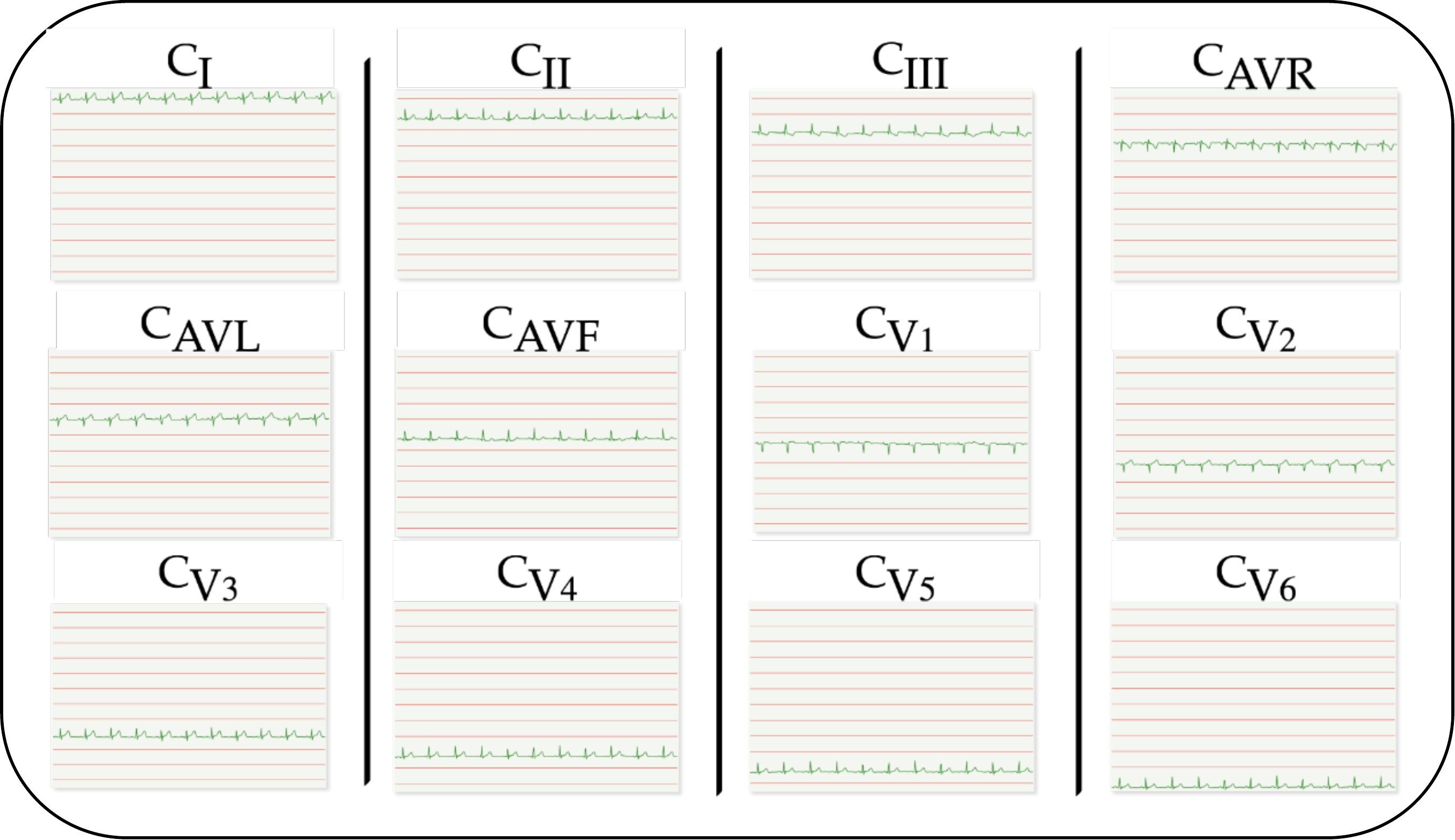}
		\caption{Masks to simulate ECG lead-reconstruction.}
		\label{fig:lead_mask}
	\end{subfigure}
	\caption{Masks applied to ECGs for simulating real-life incomplete ECGs. The green sections represent the \textit{primers} – the portions of the signal available, while the red sections indicate the parts of the signal that need to be reconstructed. In~\ref{fig:segment_mask}, the numbers in brackets specify the primers' length.}
	\label{fig:configurations}
\end{figure*}

We aimed to develop a model capable of inferring the missing portions of incomplete ECG, whether these are specific segments within leads or the entire leads themselves. To achieve this goal, we employed a data-driven approach, training a generative model on masked data to simulate real-life configurations where specific segments (or leads) of the signal are missing: in~\Cref{fig:configurations} for each of the twelve leads, the green segments denote the available segments, i.e., the \textit{primer}, while the red segments denote the part of the ECG to be reconstructed. Specifically, in~\Cref{fig:segment_mask}, we present configurations with contiguous primers varying in overall signal coverage and lead: primers of 0.8s in C1,  primers of 1.6s in C2, primers of 2.5s in C3, primers of 3.3s in C4, and primers of 5.0s in C5. In~\Cref{fig:lead_mask}, we depict configurations where the full leads must be reconstructed. The configuration the most considered in the literature is C$_{\text{I}}$, where given the full trace of the first lead, the goal is to reconstruct the remaining ones~\cite{joo2023twelve}. Although, in reality, the ECG acquisition machines record the whole signal (12 lead 10 seconds), most of this signal is lost when stored in paper ECG format or as a digital PDF file. For instance, some recording machines may capture six leads for 5 seconds, as illustrated in C5, or they might record three leads for a shorter span of 2.5 seconds, as in the case of the C3 configuration (\Cref{fig:ecg_example}). Finally, we highlight that at a higher level of abstraction, our problem definition comprehensively encompasses both ECG segment recovery and ECG lead reconstruction.

\subsection{Our proposed method}
\label{sec:ecg_recover}
Formally, let $\mathcal{S}_M\myeq\left\{\x^{(i)}\right\}_{i=1}^M\sim P_X$ be a random realization of $M$ i.i.d. valid ECG according to $P_X$. Moreover, let $\mathcal{K}$ be a countable set of indexes corresponding to each of the possible real-life data masks (e.g., as the ones we show in~\Cref{fig:configurations}). Finally, we let $\mathcal{G}\myeq\{g_{k}:\mathcal{X}\rightarrow \widetilde{\mathcal{X}}~|~ k\in\mathcal{K}\}$ be the set of possible masking functions that, given a valid ECG, outputs an incomplete one. 
Our training set is defined as
$\mathcal{D}_M \myeq\left\{\left(g(\x), \x\right)~|~ \x\in\mathcal{S}_M,~ g\in\mathcal{G}\right\}.$

Given the parametric family of models $\mathcal{F}=\left\{f_\theta:\widetilde{\mathcal{X}}\rightarrow\mathcal{X}~|~\theta\in\Theta\right\}$, where $\theta$ represents the model parameters, our learning algorithm aims to minimize the following objective function:
\begin{align}
\label{eq:main_loss}
\min_{f_\theta \in \mathcal{F}} \sum_{(\widetilde{\x}, \x) \in \mathcal{D}_m} \Big[\mathcal{L}_{\text{MSE}}( \widetilde{\x}, \x, f_\theta) + \alpha \cdot  \mathcal{L}_{\text{Pearson}}(\widetilde{\x}, \x, f_\theta)\Big],
\end{align}
where $\alpha$ is the hyper-parameter regulating the loss
and
\begin{align*}
    \mathcal{L}_{\text{MSE}}(\widetilde{\x}, \x, f_\theta)\myeq & 
    \frac{1}{L\cdot N}\sum_{l=1}^L\sum_{n=1}^N\left(f_\theta(\widetilde{\x})_{ln} - \x_{ln}\right)^2 
    \\
    \mathcal{L}_{\text{Pearson}}(\widetilde{\x}, \x, f_\theta)\myeq & \frac{1}{L}\sum_{l=1}^{L}\left(1 - r_{\x_{l}\widehat{\x}_{l}}\right),
\end{align*}
where $r_{\x_{l}\widehat{\x}_{l}}$ is the sample correlation coefficient for $(f_\theta(\widetilde{\x}), \x)\equiv(\widehat{\x}, \x)$ for the lead $l$, i.e.,
\begin{align*}
    r_{\x_{l}\widehat{\x}_{l}} = \frac{\sum_{n=1}^N\left(\widehat{\x}_{ln} - \mu_{\widehat{\x}_{l}}\right)\left(\x_{ln} - \mu_{\x_{l}}\right)}{\sqrt{\sum_{n=1}^N\left(\widehat{\x}_{ln} - \mu_{\widehat{\x}_{l}}\right)^2}\sqrt{\sum_{n=1}^N\left(\x_{ln} - \mu_{\x_{l}}\right)^2}},
\end{align*}
with $\mu_{\widehat{\x}_{l}} =\frac{1}{N}\sum_{n=1}^N\widehat{\x}_{ln}$ and $\mu_{\x_{l}}=\frac{1}{N}\sum_{n=1}^N\x_{ln}$.

The final cost function in~\Cref{eq:main_loss} accounts for the ECG's spatial and temporal aspects. Specifically, $\mathcal{L}_{\text{MSE}}$ is the mean squared error loss and ensures an accurate reconstruction of the ECG signal by capturing the average distance in amplitude between the reconstructed and the real signal. On the contrary, $\mathcal{L}_{\text{Pearson}}$ is the Pearson loss, which promotes the preservation of specific linear relationships within the signal. 
Specifically, as the imputation improves, $\mathcal{L}_{\text{Pearson}}\rightarrow 0$ as $r\rightarrow 1$ (strong correlation). 
In~\Cref{app:alpha}, we detail the performance of the model in the validation set across various values of $\alpha$, particularly focusing on the extreme cases: $\alpha=0$, i.e., when the loss is exclusively based on $\mathcal{L}_{\text{MSE}}$, and $\alpha=\infty$, i.e., the loss is $\mathcal{L}_{\text{Pearson}}$ only. Moreover, we illustrate the loss evolution during training when $\alpha=0.1$ in~\Cref{sec:loss_evolution}. 

Similarly to training, at testing starting from $T$ random realization of $U$ i.i.d. valid ECG according to $P_X$, we apply each one of the masks in $\mathcal{G}$ to create the pair of samples to test. In contrast to the training phase, our evaluation is performed separately for each set of samples generated according to the specific masks under consideration to evaluate generalization across configurations. 

\section{Results}
\label{sec:experiments}
In this section, we evaluate the effectiveness of ECGrecover and compare it with state-of-the-art (SOTA) methods. We first outline the experimental setup, and then we discuss the results. Additional simulations and findings are provided in~\Cref{app:res}  and ~\Cref{app:res2}.
\subsection{Experimental setup}
\label{sec:experimental_setup}
\subsubsection{Datasets}
\label{sec:datasets}
We utilized the Generepol dataset~\cite{salem2017genome} (cf.~\Cref{tab:database_info}), comprising 10-second 8-lead ECGs sampled at 500Hz. Leads III, aVF, aVL, and aVR were computed from leads I and II, following the common practice in the ECG literature~\cite{ECGFrontales}.

The study was conducted from 2008 to 2012 at the Clinical Investigation Center of the Pitié-Salpétrière Hospital Paris, France. 15119 ECGs were recorded from 989 healthy subjects. These recordings were taken before and after the 80mg of Sotalol oral intake. Two expert cardiologists curated and evaluated the recordings. 
We refer to~\cite{salem2017genome, prifti2021deep} for further details about the acquisition process.

The raw data for the training set consists of 4498 ECGs (i.e., $\mathcal{S}_{\text{M}}$ in~\Cref{sec:problem_formulation}), augmented to 76466 after each of the masks in~\Cref{fig:configurations} is applied. The validation set consists of 9486 ECGs, and the testing set consists of 10047 ECGs.

\subsubsection{Data preprocessing}
\label{sec:preprocessing}
We first min-max normalized the ECG data to a range of [-1,1]. We followed a strategy close to ~\cite{joo2023twelve} to prepare our data: bandpass filtering was applied to the normalized data, using lower and upper cut-off frequencies of 0.05Hz and 150Hz, respectively, to eliminate noise, such as baseline drift and improve signal quality. Finally, we downsampled the original signal from 5000pts (500Hz) to 512pts (50Hz). As described in~\Cref{sec:problem_formulation}, to mimic real-life cases of ECG with incomplete leads or segments, we considered the \textit{scenarios} illustrated in~\Cref{fig:configurations}. 
Subsequently, to maintain the original dimensions of the signal, we completed the primers with random noise. Such noise was generated from a continuous uniform distribution within the $[0, 1]$ interval. This method introduces variability in the noise across samples, thereby realistically simulating the presence of missing signals. 
An illustration of an ECG with incomplete leads or segments after introducing random noise is shown in~\Cref{fig:architecture} (top). 

\subsubsection{Evaluation metrics}
\label{sec:eval_metrics}
We assessed the effectiveness of the proposed approach from two distinct perspectives. Initially, we evaluated its reconstruction capability using standard distortion metrics employed in recent literature~\cite{joo2023twelve, seo2022multiple}. We employed the \textit{root mean square error} (RMSE) and the \textit{maximum absolute error} (MAE). Additionally, we considered the \textit{dynamic time warping} (DTW) as an alignment-based similarity measure to evaluate the distance between the original and reconstructed signals (the lower, the better) and the Pearson correlation coefficient (PCC) to evaluate the linear relationships between leads (the higher, the better). Subsequently, we examined ECGrecover's ability to preserve crucial features of the ECG signal. In particular, we focused on preserving the QT distance. We isolated each ECG heartbeat by detecting the R peak and extracting the segment 0.4 seconds before and 0.6 seconds after. For each heartbeat, we identified the position of the Q and T peaks\footnote{We used the \textit{neurokit2} python library to detect the peaks ~\cite{Makowski2021neurokit}.} to determine the QT distance. We then calculated the difference in QT distance between the reconstructed ECGs and the real ECGs denoted as $\Delta$ QT (the lower, the better). 

\subsubsection{State-of-the-art methods}
\label{sec:baselines}
We compared ECGrecover with the most recent SOTA methods for the ECG lead-reconstruction problem\footnote{No comparison could be made with Seo et al. (2022)~\cite{seo2022multiple} due to code unavailability.}. It is worth noting that, to the best of our knowledge, no other existing methods address the ECG segment-recovery problem. 

\noindent \textbf{Pix2Pix}~\cite{isola2017image}. Initially proposed for images, a conditional generative adversarial network (cGAN) employs a 2D U-Net generator and a PatchGAN discriminator. Unlike traditional discriminators that classify the entire image as real or fake, PatchGAN evaluates the realism of small, overlapping patches within the image. 

\noindent \textbf{EKGAN}~\cite{joo2023twelve}. A GAN for the reconstruction, based on Pix2Pix, augmented by an additional label generator and a 1D U-Net discriminator instead of a PatchGAN discriminator.

\noindent \textbf{CopyPaste}. A naive approach consists of replicating the primers throughout the signal to complete the ECG. This technique is implemented in~\cite{badilini2005ecgscan}, a software for digitizing paper ECGs. 

Their original paper, particularly EKGAN, focuses on reconstructing the 11 leads given lead I, corresponding to setting C$_\text{I}$ in~\Cref{fig:configurations}. Hence, the method was not initially designed for the ECG segment-recovery problem. However, for fairness, we assessed all the methods' performance across all the settings in~\Cref{fig:configurations}. 

\subsubsection{Model architecture and training procedure}
\label{sec:architecture}
\begin{figure}
    \centering
    \includegraphics[width=0.8\columnwidth]{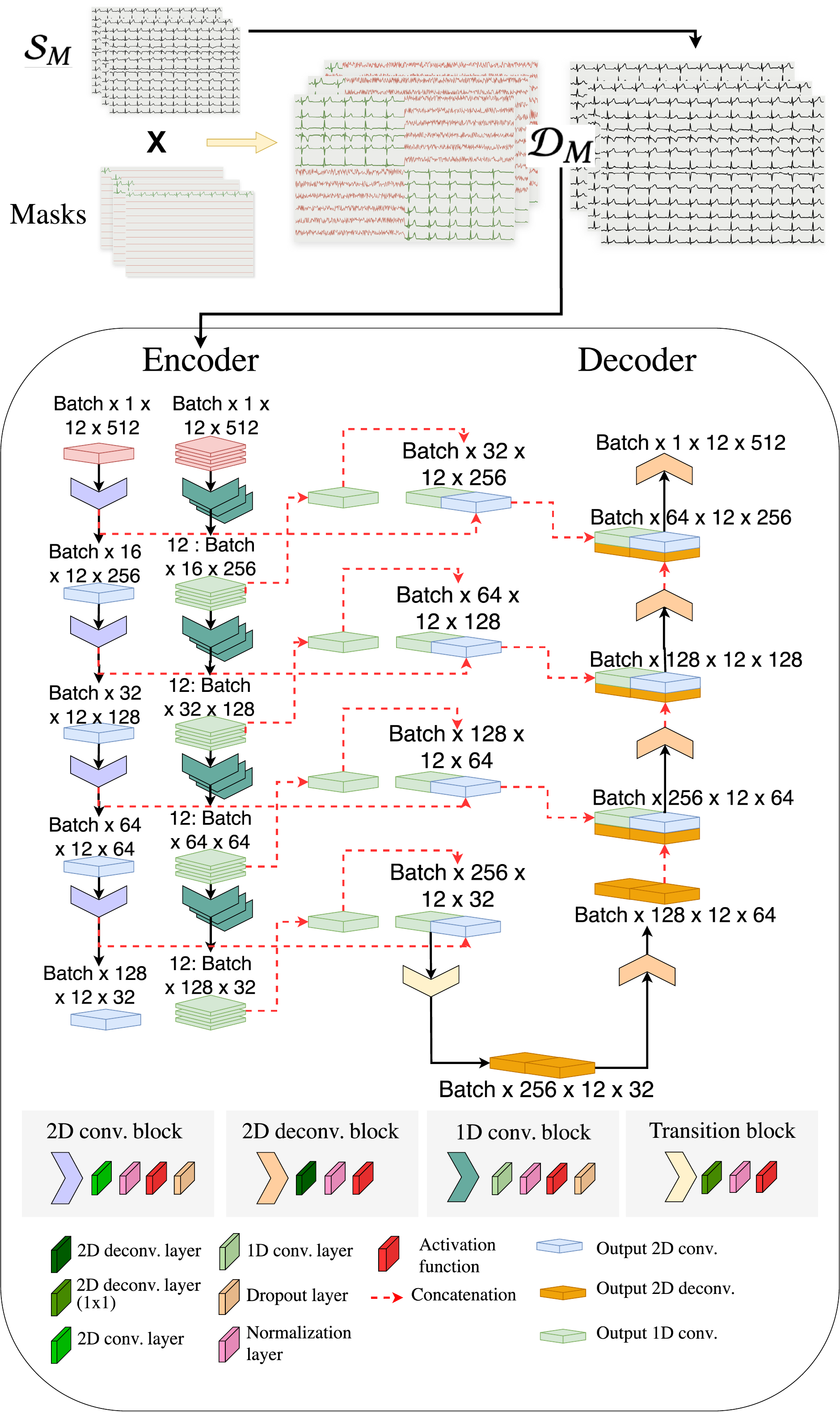}
    \caption{Training set and network architecture.}
    \label{fig:architecture}
\end{figure}
\begin{figure*}[t]
  \centering
  \includegraphics[width=0.85\textwidth]{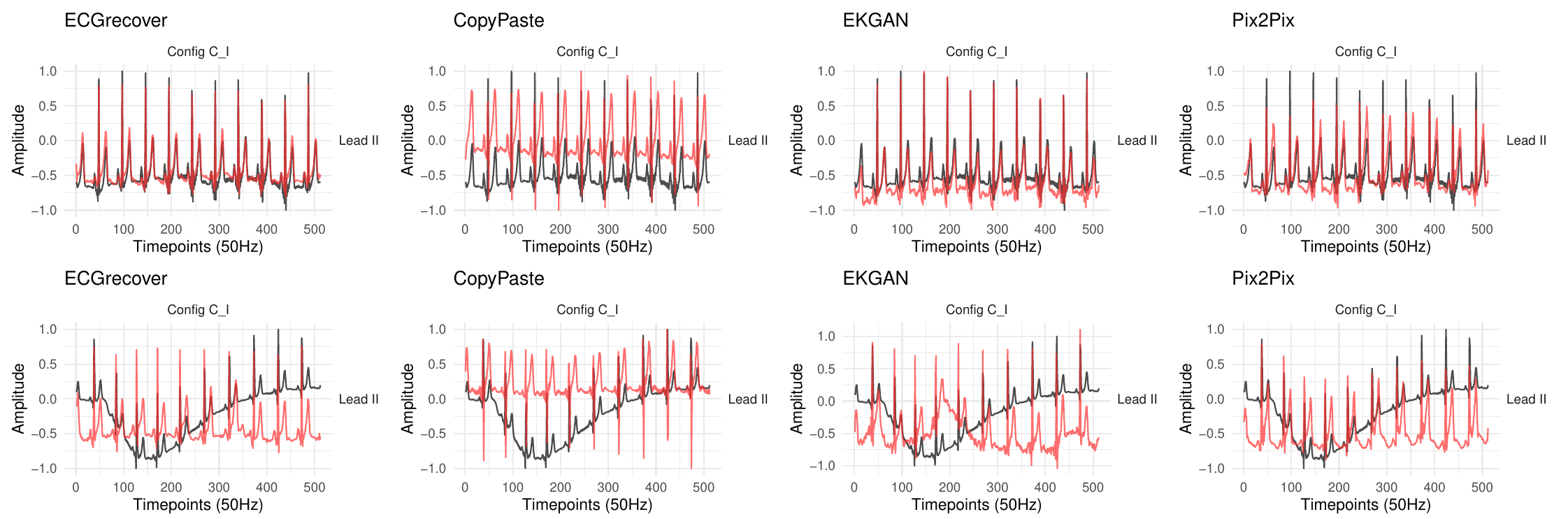}
  \caption{Lead II signal obtained with different approaches for C$_\text{I}$. Reconstructed leads are in red, the original ones are in black. Top depicts a relatively clean ECG while Bottom, an noisy ECG with baseline wander}
  \label{fig:figure_ecg_example}
\end{figure*}

We consider a U-Net like architecture~\cite{ronneberger2015u}, summarized in~\Cref{fig:architecture}. The encoder consists of two input branches, one 2D and the other 1D. On the one hand, the purpose of the 2D branch is to let the model learn inter-lead information of the signal. The 2D branch processes all leads simultaneously, facilitating the transfer of spatial information (inter-lead). This branch comprises four 2D convolution blocks, each presenting a 2D convolution layer, batch normalization, a LeakyReLu ($0.2$) activation function, and dropout ($0.2$). On the other hand, the 1D branch provides the model to learn intra-lead characteristics. The 1D branch focuses on lead-specific information, maintaining horizontal coherence (intra-lead). Similarly, to the 2D branch, this one consists of $12 \times 4$, 1D convolution blocks. The output of each block (1D/2D) is concatenated and then input in the decoder. The skip connection characteristics of the U-Net play an essential role in preserving part of the information captured by the 1D and 2D branches, which they transmit to the decoder. In this way, they maintain intra-lead and inter-lead relationships throughout the auto-encoding and decoding process.

The transition block consists of a 2D deconvolution layer, kernels of dimensions $(13,3)$ with no stride, followed by batch normalization and a LeakyReLu ($0.2$) activation function. The transition output is fed into the decoder, which consists of 4 2D deconvolution blocks. Each comprises a 2D deconvolution layer, batch normalization, and LeakyReLu activation function ($0.2$). Each output of a deconvolution block is concatenated with the output (concatenated 1D/2D) of the symmetric convolution block. Finally, in the last 2D deconvolution block of the decoder, we use the Tanh activation function to have a signal between $-1$ and $1$ as in the original normalized input signal.

Unlike SOTA approaches, our method directly tackles the challenges of segment recovery and lead reconstruction. To achieve this, it: \textit{(1)} employs a U-Net architecture that effectively captures horizontal information; \textit{(2)} offers a simpler optimization process compared to GAN-based approaches; and \textit{(3)} utilizes a hybrid 1D/2D U-Net architecture to model both inter-lead and intra-lead relationships.

We trained the model for 100 epochs with a batch size of 256. We also used the Adam optimizer with a learning rate of 0.01. The hyperparameter $\alpha$ in~\Cref{eq:main_loss} was tuned by choosing values within the range $(0,1)$. The final value of $\alpha$ used for the experiments was fixed to $0.1$ (cf.~\Cref{app:alpha}).
\subsection{Discussion}
\label{sec:disc}
We now present the main experimental results to demonstrate the effectiveness of ECGrecover in addressing lead reconstruction and segment recovery. Specifically, we first evaluate the performance of our approach using the metrics outlined in~\Cref{sec:eval_metrics}. Next, we analyze its ability to preserve key ECG features and assess the reconstructed signal's impact on inter-lead correlation. Finally, we showcase a real-world application of ECGrecover by applying it also to downstream AI tasks. Further results are provided in~\Cref{app:res}.

Throughout this section, we use the C$\bullet$ notation to indicate the input sample that, at testing time, has been masked according to the considered configuration $\bullet$. Therefore, we compare the reconstructed lead (e.g., lead I from the ECG in output when the input sample was masked with C1) with its original counterpart. 

\Cref{fig:figure_ecg_example} shows examples of lead II reconstructions. The plots suggest that ECGrecover can also correct basal drift, the effect where a signal's base axis (x-axis) appears to ‘wander’ or move up and down rather than be straight. We elaborate on this in~\Cref{app:ECG_corrected}.

\subsubsection{ECG segment-recovery and lead-reconstruction.}
\label{sec:res_recon}
\begin{figure}[t]
 \centering
 \includegraphics[width=\columnwidth]{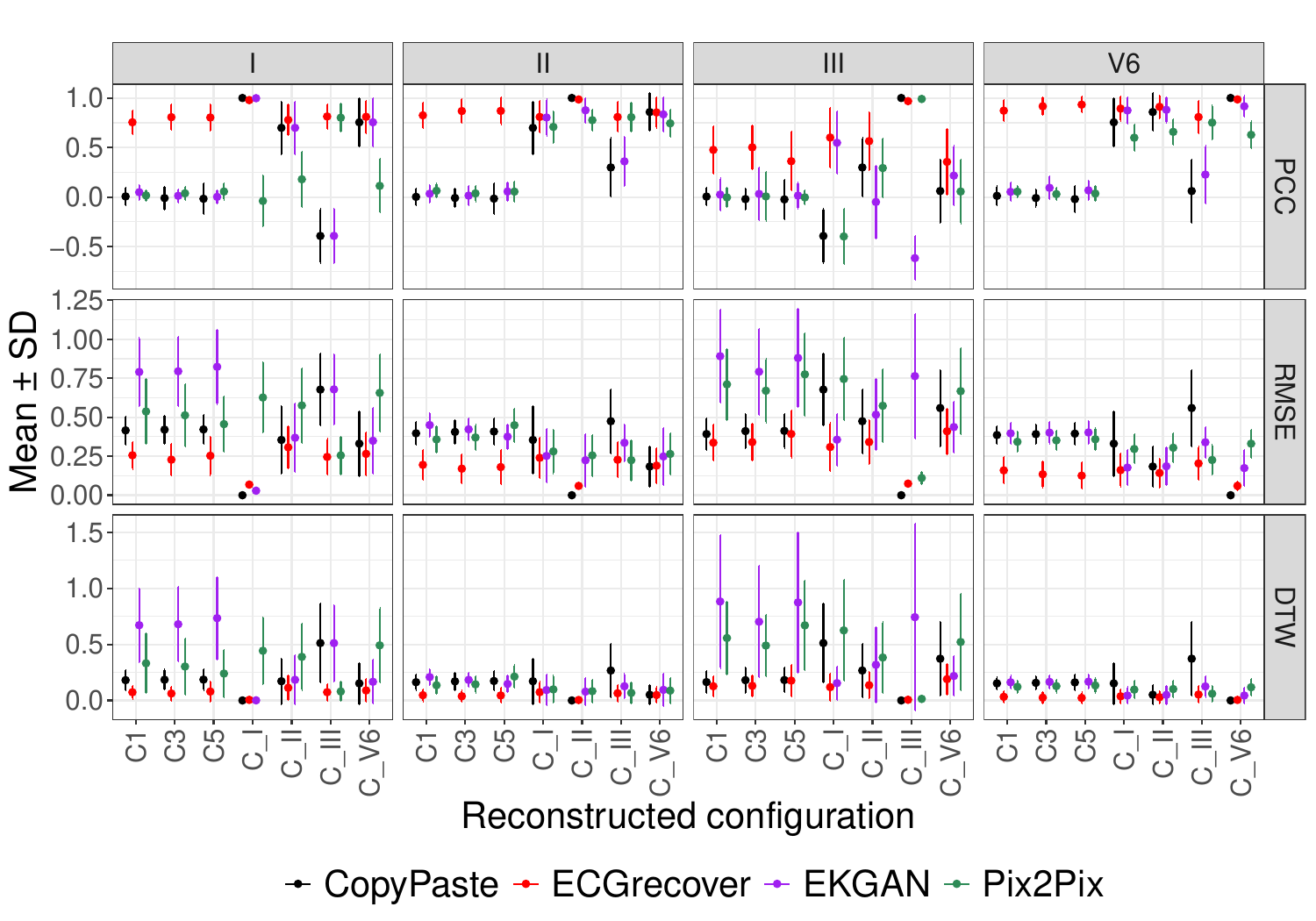}
	\caption{Performance of ECGrecover and the SOTA in their reconstruction capabilities for segment recovery (C1, C3, C5) and lead reconstruction (C$_{\text{I}}$, C$_{\text{II}}$). ECGrecover is significantly better for all metrics (p-value < 2e-16).
    }
	\label{fig:metrics}
\end{figure}
\begin{table*}[!htbp]
    \centering
    \caption{The performance of ECGRecover and the SOTA in generating coherent ECG signal, which allows to compute the QT-interval computed as the difference with that calculated in the original signal (mean$\pm$standard deviation). This is evaluated before (Sot-) and after (Sot+) drug intake, for the segment-recovery  (C1, C3, C5) and the lead-reconstruction problem (C$_{\text{I}}$, C$_{\text{II}}$).}
    \begin{tabular}{lr|cccccccc}
    \toprule
      & & \multicolumn{4}{c}{$\Delta$ $$QT distance (Sot-) $$ ($\downarrow$)} & \multicolumn{4}{c}{$\Delta$ $$QT distance (Sot+) $$ ($\downarrow$)} \\
\cmidrule(lr){1-2}\cmidrule(lr){3-6} \cmidrule(lr){7-10}
                 \multicolumn{2}{c|}{Lead}  & ECGrecover    & EKGAN~\cite{joo2023twelve}   & Pix2Pix~\cite{isola2017image} & CopyPaste  & ECGrecover    & EKGAN~\cite{joo2023twelve}     & Pix2Pix~\cite{isola2017image}  & CopyPaste\\
    \cmidrule(lr){1-2}\cmidrule(lr){3-6} \cmidrule(lr){7-10}
\multirow{3}{*}{\begin{sideways} C1\end{sideways}} &
            I & \textbf{0.01}±0.02 & 0.11±0.06 & 0.09±0.04 & 0.05±0.05 
          & \textbf{0.02}±0.03 & 0.12±0.07 & 0.11±0.05 & 0.07±0.05  \\
          & II & \textbf{0.01}±0.01 & 0.12±0.06 & 0.08±0.04 & 0.04±0.04  
          & \textbf{0.01}±0.03 & 0.15±0.07 & \textbf{0.1}±0.05 & 0.06±0.07 \\
          & III & 0.07±0.07 & 0.11±0.09 & 0.11±0.09 & \textbf{0.05}±0.08 
          & \textbf{0.06}±0.06 & 0.11±0.09 & 0.13±0.09 & \textbf{0.06}±0.07  \\
          & V1 & 0.03±0.04 & 0.05±0.05 & 0.04±0.04 & \textbf{0.02}±0.04
          & 0.03±0.04 & 0.05±0.05 & 0.04±0.04 & \textbf{0.02}±0.04  \\ 
          & V6 & \textbf{0.01}±0.01 & 0.13±0.06 & 0.06±0.04 & 0.02±0.05 
          & \textbf{0.01}±0.03 & 0.16±0.06 & 0.08±0.06 & 0.03±0.06  \\

\cmidrule(lr){1-2} \cmidrule(lr){3-6} \cmidrule(lr){7-10}
\multirow{3}{*}{\begin{sideways} C3\end{sideways}} &
            I & \textbf{0.01}±0.01 & 0.05±0.04 & 0.08±0.04 & 0.02±0.03 
            & \textbf{0.01}±0.03 & 0.07±0.05 & 0.09±0.05 & 0.03±0.04\\
          & II & \textbf{0.01}±0.01 & 0.07±0.05 & 0.05±0.04 & 0.02±0.02
          & \textbf{0.01}±0.02 & 0.10±0.06 & 0.07±0.04 & 0.03±0.04  \\
          & III & 0.06±0.07 & 0.11±0.09 & 0.10±0.08 & \textbf{0.04}±0.06
          & \textbf{0.05}±0.05 & 0.11±0.09 & 0.13±0.09 & \textbf{0.05}±0.06 \\
          & V1 & 0.03±0.03 & 0.04±0.04 & 0.04±0.04 & \textbf{0.02}±0.03 
          & \textbf{0.02}±0.04 & 0.04±0.05 & 0.03±0.04 & \textbf{0.02}±0.02\\ 
          & V6 & \textbf{0.01}±0.01 & 0.10±0.05 & 0.05±0.04 & 0.02±0.03 
          & \textbf{0.01}±0.02 & 0.13±0.06 & 0.07±0.05 & 0.03±0.04 \\

\cmidrule(lr){1-2} \cmidrule(lr){3-6} \cmidrule(lr){7-10}
\multirow{3}{*}{\begin{sideways} C5\end{sideways}} &
            I & \textbf{0.01}±0.02 & 0.03±0.02 & 0.03±0.03 & \textbf{0.01}±0.02 
            & \textbf{0.01}±0.03 & 0.04±0.04 & 0.05±0.04 & 0.02±0.02\\
          & II & \textbf{0.01}±0.01 & 0.04±0.04 & 0.03±0.03 & \textbf{0.01}±0.02 
          & \textbf{0.01}±0.02 & 0.07±0.06 & 0.04±0.04 & \textbf{0.01}±0.02 \\
          & III & 0.05±0.06 & 0.09±0.08 & 0.08±0.08 & \textbf{0.02}±0.04 
          & 0.05±0.06 & 0.12±0.09 & 0.10±0.09 & \textbf{0.02}±0.04 \\
          & V1 & 0.03±0.03 & 0.04±0.04 & 0.05±0.04 & \textbf{0.01}±0.02
          & 0.02±0.03 & 0.03±0.05 & 0.05±0.04 & \textbf{0.01}±0.02\\ 
          & V6 & \textbf{0.00±}0.01 & 0.05±0.04 & 0.02±0.02 & 0.01±0.01 
          & \textbf{0.01}±0.02 & 0.09±0.06 & 0.04±0.04 & 0.02±0.02 \\
\cmidrule(lr){1-2} \cmidrule(lr){3-6} \cmidrule(lr){7-10}
\multirow{3}{*}{\begin{sideways} C$_{\text{I}}$\end{sideways}} &
           II & \textbf{0.01}±0.01 & \textbf{0.01}±0.01 & \textbf{0.01}±0.03 & \textbf{0.01}±0.02 
          & \textbf{0.01}±0.03 & \textbf{0.01}±0.03 & 0.02±0.04 & \textbf{0.01}±0.03  \\
          & III & \textbf{0.06}±0.07 & 0.08±0.08 & 0.09±0.08 & 0.08±0.07
          & \textbf{0.06}±0.07 & 0.07±0.07 & 0.09±0.07 & 0.09±0.07\\
          & V1 & \textbf{0.03}±0.04 & 0.04±0.04 & 0.10±0.05 & 0.09±0.06 
          & \textbf{0.03}±0.04 & \textbf{0.03}±0.04 & 0.12±0.06 & 0.12±0.06\\ 
          & V6 & \textbf{0.01}±0.01 & \textbf{0.01}±0.01 & 0.02±0.02 & \textbf{0.01}±0.02
          & \textbf{0.01}±0.03 & \textbf{0.01}±0.03 & 0.03±0.03 & \textbf{0.01}±0.03 \\

\cmidrule(lr){1-2} \cmidrule(lr){3-6} \cmidrule(lr){7-10}
\multirow{3}{*}{\begin{sideways} C$_{\text{II}}$\end{sideways}} &
           I & \textbf{0.01}±0.02 & \textbf{0.01}±0.02 & 0.02±0.02 & \textbf{0.01}±0.02
          & \textbf{0.01}±0.04 & \textbf{0.01}±0.04 & 0.03±0.04 & \textbf{0.01}±0.04 \\
          & III & 0.09±0.10 & 0.09±0.08 & \textbf{0.08}±0.07 & \textbf{0.08}±0.06
          & \textbf{0.08}±0.09 & 0.11±0.08 & \textbf{0.08}±0.07 & 0.09±0.07\\
          & V1 & \textbf{0.03}±0.04 & 0.04±0.04 & 0.09±0.05 & 0.09±0.05 
          & \textbf{0.03}±0.04 & \textbf{0.03}±0.04 & 0.09±0.06 & 0.12±0.06\\ 
          & V6 & \textbf{0.01}±0.01 & \textbf{0.01}±0.01 & 0.02±0.02 & \textbf{0.01}±0.01 
          & \textbf{0.01}±0.02 & \textbf{0.01}±0.03 & 0.02±0.03 & \textbf{0.01}±0.03 \\

    \bottomrule
    \end{tabular}
    \label{tab:peak_rmy_label}
\end{table*}

\Cref{fig:metrics} presents an evaluation of ECGrecover, alongside benchmark state-of-the-art methods, focusing on PCC, RMSE and DTW for the Generepol dataset. We relegate to~\Cref{fig:metrics_2_dataset} the complete sets of the results across all the 17 \textit{scenarios} and all the 12 leads. 

As for the ECG segment-recovery problem, in~\Cref{fig:metrics}, we specifically focused on configuration C1 as it represents the scenario with primers of shorter duration (i.e., 0.8sec) where most of the signal needs to be reconstructed. Additionally, we analyze the settings C3 and C5 as they reflect real-life situations when digitizing paper ECGs. Regarding the ECG lead-reconstruction problem, we highlight in the figure the configurations C$_\text{I}$ as it is the one considered also considered in~\cite{joo2023twelve}, and C$_\text{II}$ 
because of its relevance in wearable device recordings and paper-stored ECGs, where Lead II, akin to Lead I, is predominantly captured. For each of the input masks, we tested the ability of the model to reconstruct the limb leads I, II, and III and the chest leads V1 and V6. 
Practitioners note that leads II and V1 are useful leads for measuring heart rate.

Overall, ECGrecover outperforms or reaches the same results as the SOTA, regardless of the chosen metric. As depicted in~\Cref{fig:metrics}, lead V6 emerges as the most straightforward to reconstruct, both in ECG segment-recovery (C1, C3, C5) and ECG lead-reconstruction (C${\text{I}}$, C${\text{II}}$). Conversely, lead III is the most challenging to reconstruct from the other leads. In particular, while our method consistently delivers satisfactory results across all samples tested, competing methods exhibit more significant variability (e.g., DTW values for EKGAN). Interestingly, the simple CopyPaste method yields reasonable results regarding RMSE and DTW. However, as seen from the PCC results, the method fails to preserve the linear relationships between the original and reconstructed leads, underscoring the method's limitations as the heart rhythm may vary through time, thus losing sync. This highlights the importance of evaluating the methods from multiple perspectives. 

Referring to the comprehensive results in~\Cref{fig:metrics_2_dataset}, our analysis shows that, in contrast to ECGrecover, the competing approaches EKGAN and Pix2Pix face significantly more challenges in the ECG segment-recovery task compared to the ECG lead-reconstruction task. On the other hand, the CopyPaste method struggles primarily with the lead-reconstruction problem. This outcome is not surprising as EKGAN and Pix2Pix were specifically designed for ECG lead-reconstruction. In contrast, CopyPaste, which essentially duplicates the signal across leads, performs poorly when the lead to reconstruct is less correlated with the primer lead in the mask. We emphasize that in~\Cref{app:rst_ptb}, we conduct an additional evaluation of our proposed method using the PTB-XL dataset~\cite{wagner2020ptb}. 

\subsubsection{On the QT distance preservation}
\label{sec:res_peaks}
An ECG visually represents the heart's electrical activity during a cardiac cycle, encompassing a single heartbeat, which involves the heart filling with blood and then contracting to pump it out. The standard ECG trace includes several key components, each corresponding to specific electrical events within the heart. The intervals and segments between these components (QT distance, PR interval, etc.) also provide important information about the cardiac cycle's timing and electrical conduction through the heart. Understanding and evaluating them helps diagnose a wide range of cardiac abnormalities and conditions~\cite{parmet2003electrocardiograms}. In~\Cref{tab:peak_rmy_label}, we evaluated the methods' capacity to generate a signal that can be effectively analyzed. The performance is computed as the difference between the QT distance calculated in the reconstructed signals and the QT distance computed in the original signals ($\Delta$-QT). Specifically, we segmented each ECG into heart-beats by identifying the position of the R-peak within the original leads and isolated the specific spike by taking a 1-second window centered on the R-peak. For detecting the peaks, we use the Neurokit library~\cite{Makowski2021neurokit}, which supports extracting critical features from the ECG. At each heartbeat, we identified the Q wave and the end of the T wave, and calculate the average QT interval for each ECG. 

The results in~\Cref{tab:peak_rmy_label} are grouped w.r.t. Sot+ and Sot- true label in the Generepol study. Indeed, as described in~\Cref{sec:datasets}, in this study participants were administered Sotalol, a drug known to prolong the QT interval and in some cases alter the morphology of the T-wave. These effects make QT interval measurement more challenging~\cite{salem2017genome, prifti2021deep}. Remarkably, our analysis did not reveal a significant performance difference in ECGrecover between ECGs recorded before and after drug intake. The method demonstrate consistent performance across various input masks and lead reconstructions. While ECGrecover outperforms or reaches the same results as its competitors, the CopyPaste method occasionally achieves better results. The explanation lies in the inherent nature of the CopyPaste technique, which maintains beat information through direct replication. By duplicating each beat identically, the CopyPaste method inherently preserves the average QT distance with lower variability, reflecting an artifact. 

\subsubsection{On the inter-lead correlation.}
\label{sec:res_lead}
\begin{figure}[!htbp]
    \centering
    \includegraphics[width=.8\columnwidth]{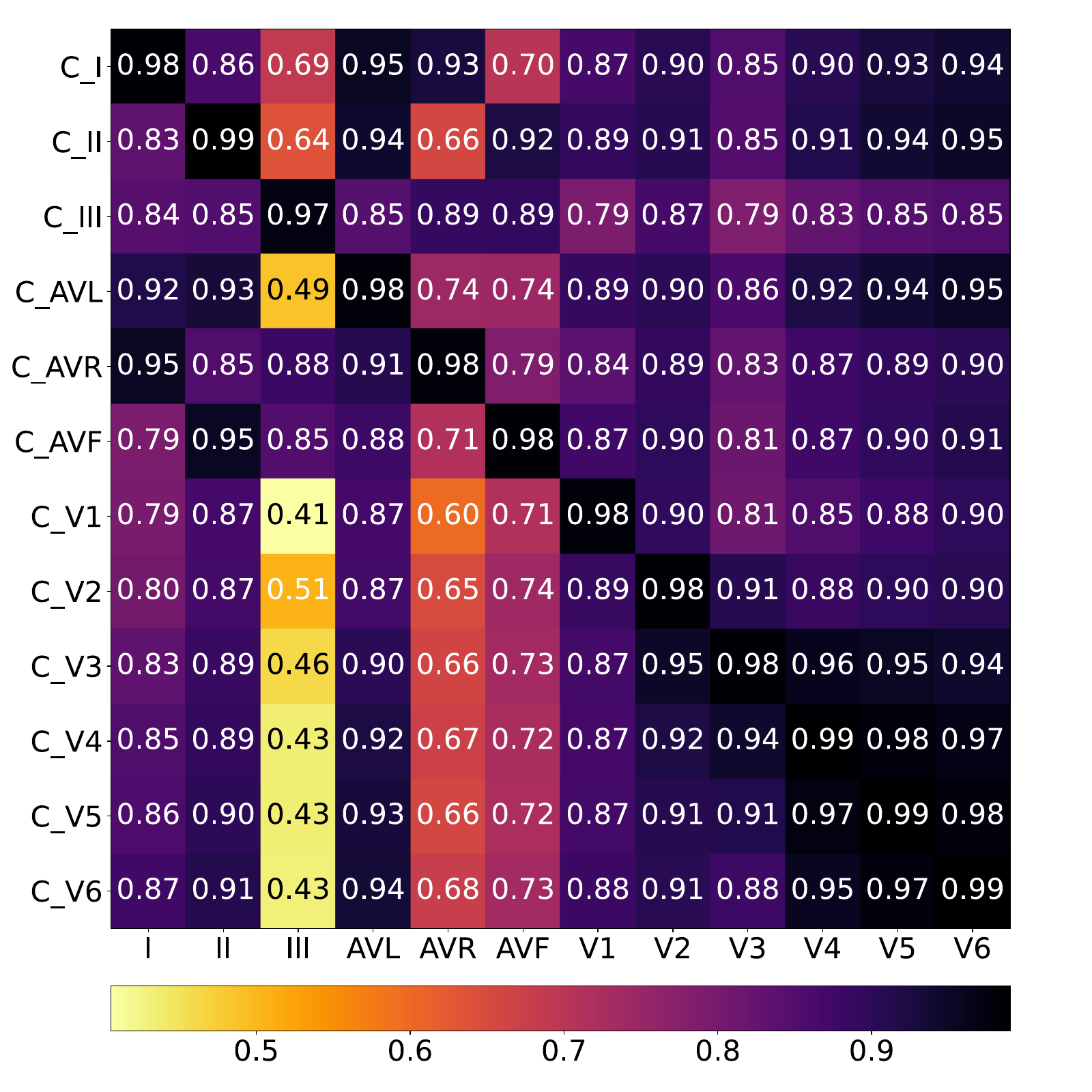}
    \caption{
    The matrix entry at index $(i, j)$ represents PCC between the $j$-th reconstructed lead and the $j$-th original lead, given that the input mask applied is $C_{i}$.
    }
    \label{fig:Correlation_matrix}
\end{figure}
We are interested in inter-lead correlation during ECG reconstruction. Specifically, in~\Cref{fig:Correlation_matrix}, the matrix entry at index $(i, j)$ represents PCC between the $j$-th reconstructed lead and the $j$-th original lead, given that the input mask applied is $C_{i}$. Thus, the $i$-th row illustrates the model's effectiveness in utilizing the specific $i$-th lead's data to reconstruct the remaining leads of the ECG accurately. Conversely, the $j$-th column indicates which input mask most significantly guides the model towards a more accurate reconstruction.

\Cref{fig:Correlation_matrix} highlights the strong correlation between leads V3, V4, V5 and V6. Lead III exhibits a lower correlation with the other leads. While surprising at first, given that lead III can theoretically be derived from leads I and II through Einthoven's triangle~\cite{parmet2003electrocardiograms}, this characteristic has also been observed in~\cite{jain2023redundancy} on a distinct dataset. Moreover, similar conclusions can also be found in~\cite{zhang202212} regarding lead V1. Their findings revealed a notably low correlation between lead V1 and the other leads, with PCC close to 0, suggesting that the information provided by lead V1 is substantially distinct from that of the other leads. Consequently, it is understandable that the mask C$_{\text{V1}}$ in~\Cref{fig:Correlation_matrix} is the most challenging to reconstruct. Noteworthy, PCC measures linear correlation, and other complex relationships will not be captured. Factors like noise or artifacts in individual physiology can also contribute to lower correlation values.

\subsubsection{Reconstructing digitized paper-stored ECGs}
\label{app:real_life_app}
\begin{table*}[t]
    \centering
    \caption{ECGrecover for reconstructing digitized paper-based ECGs. C$_{\text{real-life}}$ corresponds to C3 configuration with lead II fully represented. The values are computed by comparing the numerical signal provided in the datasets with the signal obtained after reconstructing the digitized paper-stored ECG.}
    \begin{tabular}{lr|ccc|ccc}
    \toprule
      & & \multicolumn{3}{c}{PTB-XL (2203 ECGs)}   & \multicolumn{3}{|c}{LUDB (200 ECGs)}\\
\cmidrule(lr){1-2}\cmidrule(lr){3-5}\cmidrule(lr){6-8}
                 \multicolumn{2}{c|}{Lead}  & PCC ($\uparrow$)    & RMSE ($\downarrow$)   & DTW ($\downarrow$) & PCC ($\uparrow$)    & RMSE ($\downarrow$)  & DTW ($\downarrow$)\\
                 
\cmidrule(lr){1-2}\cmidrule(lr){3-5}\cmidrule(lr){6-8}

\multirow{12}{*}{\begin{sideways} C$_\text{real-life}$ \end{sideways}} &
           I & 0.96±0.18 & 0.04±0.04 & 0.0±0.0 
           & 0.96±0.20 & 0.03±0.02 & 0.0±0.0\\
          & II & 0.95±0.18 & 0.05±0.06 & 0.0±0.0 
          & 0.95±0.21  &	0.04±0.04 &	0.0±0.0\\
          & III & 0.92±0.21 & 0.05±0.06 & 0.0±0.0 
          & 0.91±0.22 &	0.04±0.03 &	0.0±0.0\\
          & AVL & 0.91±0.23 & 0.05±0.09 & 0.0±0.0
          & 0.91±0.24 &	0.06±0.07 &	0.0±0.0\\ 
          & AVR & 0.95±0.21 & 0.05±0.06 & 0.0±0.0
          & 0.95±0.22 &	0.04±0.04 &	0.0±0.0\\
          & AVF & 0.93±0.21 & 0.05±0.08 & 0.0±0.0 
          & 0.92±0.21 &	0.05±0.05 &	0.0±0.0\\
          & V1 & 0.86±0.31 & 0.09±0.23 & 0.0±0.0 
          & 0.86±0.31 &	0.07±0.09 &	0.0±0.0\\
          & V2 & 0.90±0.32 & 0.17±0.27 & 0.0±0.0 
          & 0.87±0.36 &	0.08±0.09&	0.0±0.0\\
          & V3 & 0.87±0.31 & 0.27±0.35 & 0.0±0.0
          & 0.87±0.30 &	0.13±0.11 &	0.0±0.0\\
          & V4 & 0.90±0.29 & 0.21±0.32 & 0.0±0.0
          & 0.88±0.31 &	0.09±0.11 &	0.0±0.0\\
          & V5 & 0.93±0.28 & 0.13±0.23 & 0.0±0.0 
          & 0.92±0.32 &	0.07±0.09 &	0.0±0.0\\
          & V6 & 0.94±0.28 & 0.09±0.28 & 0.0±0.0 
          & 0.94±0.30 &	0.06±0.07 &	0.0±0.0\\
    \bottomrule
    \end{tabular}
    \label{tab:metric_real_life}
\end{table*}
Historically, ECGs have been printed and stored on paper or saved as images~\cite{Einthoven1903}, which presents clear storage and exploitation challenges, as highlighted in~\Cref{sec:introduction}. In the most common formats for paper-stored ECGs, each lead is only partially represented, typically showing only 2.5 or 5 seconds of the usual 10-second signal. This section evaluates the proposed method in a real-life scenario by examining ECGrecover's ability to reconstruct the missing parts of ECGs that were initially stored on paper and digitized.

We considered the LUDB~\cite{kalyakulina2020ludb} and PTB-XL~\cite{wagner2020ptb} datasets for the experiments. Specifically, these datasets are available in both the original paper-stored version (in \texttt{.pdf}) and the corresponding numerical version (in \texttt{.xml}). 
The ECGs saved in \texttt{.pdf} format was digitized with ECGminer software~\cite{santamonica2024ecgminer} to produce a partial \texttt{.xml} version. The \texttt{.pdf} version of the ECGs corresponds to configuration C3, except for lead II, which is complete. We refer to this new setting as C$_{\text{real-life}}$. In contrast, the \texttt{.xml} version includes the full 12-lead tracks, making it well-suited for comprehensive testing of the reconstruction method. We evaluated ECGrecover, trained on the Generepol dataset, to assess its robustness in reconstructing ECGs from various databases. To align with the training phase, the signals were preprocessed in the same way as presented in ~\Cref{sec:preprocessing}.

The results presented in~\Cref{tab:metric_real_life} demonstrate that the proposed method achieves strong performance in this setting. Specifically, for the PTB-XL database, the method yields an overall mean PCC of 0.92, an RMSE of 0.1, and an SDTW of 0. For the LUDB database, it achieves an average PCC of 0.91, an RMSE of 0.06, and a DTW distance of 0. These positive outcomes can be attributed to the carefully designed configuration we sought to reconstruct. In particular, C${\text{real-life}}$ incorporates the full lead II along with partial versions of other leads. This configuration serves as an intermediate setup between C${\text{3}}$ and C$_{\text{II}}$, providing the model with a richer and more diverse set of input information compared to the training configurations. The enhanced input likely accounts for the improved predictive performance. Notably, as illustrated in~\Cref{fig:Correlation_matrix}, lead II alone demonstrates the ability to reconstruct all ECG leads with an accuracy of 0.87. By integrating additional information from other leads, the hybrid configuration in C$_{\text{real-life}}$ likely amplifies the model’s performance, resulting in the observed improvements.

\subsubsection{Applying ECGrecover to AI-based applications}
Here we illustrate the use of ECGrecover in a downstream AI-based application task, predicting the risk of Torsade-de-Pointes events. As detailed in~\Cref{sec:datasets}, Sot- and Sot+ denote ECGs from healthy individuals before and after ingesting 80mg of Sotalol. Thus, we utilize the CNN model initially crafted by Prifti et al.~\cite{prifti2021deep} for TdP risk prediction. This DenseNet model comprises six blocks, each containing eight dense convolutional layers. We trained it with the Adam optimizer, employing a learning rate of 0.001 and a dropout rate of 0.2. We evaluate our method as a data augmentation technique.

During training, for a given input sample $\x$, we augmented the dataset with $\x$, recovered from configuration C$\cdot$, with a 0.25 probability and trained the TdP-risk model for 50 epochs, iterating the training phase five times with seeds [0-4] for the random number generator. We focused on configurations C3 and C5, commonly encountered in real-life settings, as discussed in~\Cref{sec:res_recon}. The TdP-risk model's accuracy on the original testing set without data augmentation is 92.47\%. In contrast, with data augmentation, the accuracy is 93.84\%±0.0065 for C3, 93.73\%±0.002 for C5, and 92.17\%±0.0128 for C$_{\text{I}}$.
Results are presented as mean +/- standard deviation across the five training stages.
The results show that data regenerated with ECGrecover can be used to augement the initial dataset, thus improving model performance. In particular, data regenerated from C3 increases model performance by 1\%.
\section{Conclusions and future work}
We tackled the problem of reconstructing the 12-lead ECG signal when only a partial subset of the signal is available. We proposed a U-Net-like model trained on a novel objective function, combining the mean squared error loss with the Pearson correlation. We evaluated ECGrecover on a real-life ECG dataset comprising 15 119 10-second ECGs. The comparison to recent SOTA revealed the effectiveness of ECGrecover in recovering the missing signal, even in challenging cases such as the C$_{\text{I}}$ configuration, as quantified either with distortion metrics or by the preservation of the QT segment. 

By using the reconstruction capabilities of ECGrecover, we can improve the readability of ECG traces, which can be determinant in diagnosing diseases. This advancement allows for a more detailed and accurate assessment of cardiac activity, facilitating better diagnostic decisions. Notably, ECGrecover is part of the DeepECG4U\footnote{\url{https://anr.fr/Project-ANR-20-CE17-0022}} translational research project, which aims to improve patient care through automatic analyses of ECG based on AI.


\section*{Acknowledgements}
This study was supported by the ANR-20-CE17-0022 DeepECG4U funding from the French National Research Agency (ANR).
\printbibliography 

\appendix
\section{Extended analysis of ECGrecover}
\label{app:res}
This section presents additional analysis that could not be included in the main paper due to space constraints. 
\subsection{On the $\alpha$ parameter}
\label{app:alpha}
In~\Cref{sec:ecg_recover}, we introduced the objective function in~\Cref{eq:main_loss} used to train our U-Net model:

$$\mathcal{L}=\mathcal{L}_{\text{MSE}} + \alpha \cdot \mathcal{L}_{\text{Pearson}}.$$
In the following section, we investigate the hyperparameter $\ alpha$'s role in training the model. In particular, we consider $\alpha\in\{0, 0.1, 0.5, 1, \infty\}$, where for $\alpha=0$, $\mathcal{L}\equiv\mathcal{L}_{\text{MSE}}$ and for $\alpha=\infty$, $\mathcal{L}\equiv\mathcal{L}_{\text{Pearson}}$. We provide in~\Cref{tab:table_alpha} the results over the validation set of the five models trained with the different values of $\alpha$.
As we can see, when the model is trained on the mean squared error alone, it focuses solely on minimizing the magnitude differences between the predicted and original ECG signals. Notably, while the results for RMSE and MAE remain consistent for the models with $\alpha\leq 1$, the values for the PCC and DTW present higher variability, with an improvement of 4.65\% in terms of PCC for the model with $\alpha=0.5$ w.r.t. the model with $\alpha=0$ and of 10.44\% in terms of DTW. Interestingly, setting $\alpha=\infty$, resulted in the model performing well in terms of PCC but significantly worse across other metrics. This observation underscores the importance of employing a hybrid loss considering the ECG signal's spatial and temporal aspects. 

For our testing phase, we chose 
$\alpha=0.1$ based on its performance during validation, showing promising results compared to $\alpha=0$ and $\alpha=1$, without extensive exploration of other values. However, a more detailed examination revealed 
$\alpha=0.5$ to potentially offer better outcomes. This finding suggests that while 
$\alpha=0.1$ provided satisfactory results, a further optimization of $\alpha$ could enhance model performance. In future work, we will investigate this aspect more thoroughly to identify the most effective $\alpha$ parameter value.
\begin{table}[!htbp]
\caption{ECGrecover performances for different values of $\alpha$ in~\Cref{eq:main_loss}. We note $\infty$ when we consider only the PCC}
\begin{tabular}{l|cccc}
\toprule
$\alpha$  & PCC & RMSE & MAE  & DTW \\
\midrule
0        & 0.86        & 0.19 & 0.15 & 18.10    \\
0.1      & 0.88        & \textbf{0.18} & \textbf{0.14} & 16.87    \\
0.5      & 0.90        & \textbf{0.18} & \textbf{0.14} & \textbf{16.21}    \\
1        & 0.90        & \textbf{0.18} & 0.15 & 16.52    \\
$\infty $   & \textbf{0.91}     & 0.43 & 0.39 & 95.37    \\
\bottomrule
\end{tabular}
\label{tab:table_alpha}
\end{table}

\subsection{Loss evolution during training ($\alpha=0.1$)}
\label{sec:loss_evolution}
\begin{figure}[!htbp]
    \centering
    \includegraphics[width=0.95\columnwidth]{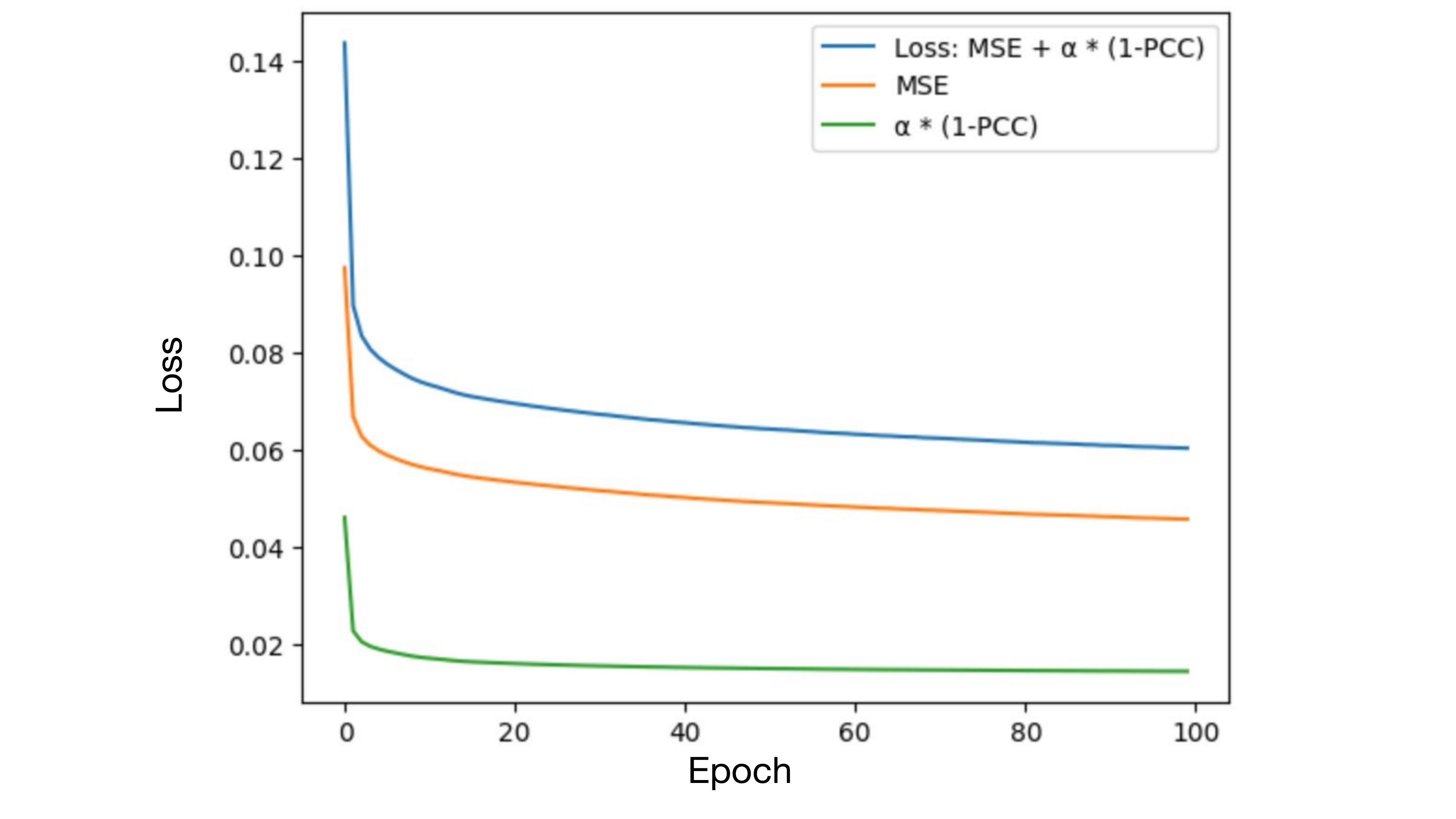}
    \caption{
    Evolution of the loss components during training
    }
    \label{fig:FigLoss}
\end{figure}

\Cref{fig:FigLoss} illustrates the progression of the total loss (in blue), $\mathcal{L}_{\text{MSE}}$ (in orange), and $\mathcal{L}_{\text{Pearson}}$  (in green) across 100 epochs. The total loss decreased, driven by the decline of the MSE and Pearson terms. The rapid reduction of the Pearson term during the early epochs indicates that the network quickly learns to align its predictions with the desired correlation structure. Meanwhile, the gradual decrease in MSE reflects ongoing improvements in reconstruction accuracy.

\subsection{Impact of ECGrecover on signal quality assessment}
\label{app:ECG_corrected}
\begin{table}[!htbp]
    \footnotesize
    \centering
    \caption{
Comparison of the average quality scores between the original ECGs in the Generepol test set and those reconstructed using ECGrecover. 
The SQI${\text{Avg-QRS}}$~\cite{Makowski2021neurokit} metric assigns a numerical quality score, where higher values indicate better ECG quality. The SQI${\text{Fuzzy}}$~\cite{zhao2018sqi} method offers a qualitative assessment as a string. For this metric, we report the percentage of \textit{unacceptable} ECGs.}
\begin{tabular}{>{\centering\arraybackslash}p{10mm}|cccc}
\toprule
 \multicolumn{1}{c}{} & \multicolumn{2}{c}{SQI$_{\text{Avg-QRS}}$ ($\uparrow$)}   & \multicolumn{2}{c}{SQI$_{\text{Fuzzy}}$ \% ($\downarrow$)  }\\
 \midrule
                Config.  & Original & ECGrecover & Original  & ECGrecover \\
                 
\cmidrule(lr){1-1} \cmidrule(lr){2-3} \cmidrule(lr){4-5}

           C1& \textbf{0.52}$\pm$0.09 & \textbf{0.52}$\pm$0.08 & 6$\pm$23& \textbf{3}$\pm$17  \\
            C2& \textbf{0.52}$\pm$0.09 & 0.51$\pm$0.09 & 6$\pm$23& \textbf{3}$\pm$17   \\
            C3& \textbf{0.52}$\pm$0.09 & 0.51$\pm$0.09 & 6$\pm$23& \textbf{3}$\pm$16\\
            C4 & \textbf{0.52}$\pm$0.09 & \textbf{0.52}$\pm$0.09 & 6$\pm$23& \textbf{3}$\pm$18  \\ 
           C5 & \textbf{0.52}$\pm$0.09 & 0.48$\pm$0.10 & 6$\pm$23& \textbf{3}$\pm$18  \\
          \cmidrule(lr){1-1} \cmidrule(lr){2-5} 
           C$_{\text{I}}$ & 0.52$\pm$0.09 & \textbf{0.53}$\pm$0.09 & 6$\pm$23& \textbf{3}$\pm$18 \\
           C$_{\text{II}}$ & 0.52$\pm$0.09 & \textbf{0.53}$\pm$0.08 & 6$\pm$23& \textbf{2}$\pm$13\\
           C$_{\text{III}}$ & 0.52$\pm$0.09 & \textbf{0.53}$\pm$0.08 & \textbf{6}$\pm$23& 16$\pm$36 \\
           C$_{\text{AVL}}$ & 0.52$\pm$0.09 & \textbf{0.53}$\pm$0.08 & 6$\pm$23& \textbf{1}$\pm$11 \\
          C$_{\text{AVR}}$ & 0.52$\pm$0.09 & \textbf{0.53}$\pm$0.09 & \textbf{6}$\pm$23& \textbf{6}$\pm$24  \\
           C$_{\text{AVF}}$ & 0.52$\pm$0.09 & \textbf{0.53}$\pm$0.09 & 6$\pm$23 & \textbf{3}$\pm$18  \\
           C$_{\text{V1}}$ & 0.52$\pm$0.09 & \textbf{0.53}$\pm$0.09 & 6$\pm$23& \textbf{2}$\pm$15  \\
           C$_{\text{V2}}$ & 0.52$\pm$0.09 & \textbf{0.53}$\pm$0.08 & 6$\pm$23& \textbf{1}$\pm$8  \\
          C$_{\text{V3}}$ & 0.52$\pm$0.09 & \textbf{0.53}$\pm$0.08 & 6$\pm$23& \textbf{2}$\pm$15  \\
          C$_{\text{V4}}$ & 0.52$\pm$0.09 & \textbf{0.53}$\pm$0.08 & 6$\pm$23& \textbf{1}$\pm$12 \\
           C$_{\text{V5}}$ & 0.52$\pm$0.09 & \textbf{0.53}$\pm$0.08 & 6$\pm$23& \textbf{1}$\pm$9  \\
           C$_{\text{V6}}$ & 0.52$\pm$0.09 & \textbf{0.53}$\pm$0.08 & 6$\pm$23& \textbf{1}$\pm$8  \\
    \bottomrule
    \end{tabular}
    \label{tab:quality_check}
\end{table}

                 


In this section, we examine the impact of ECGrecover's reconstruction on ECG quality. ECG recordings are frequently susceptible to noise from various sources, including loose electrode contacts, patient movement, and muscle contractions. These factors can distort the waveform, complicating analysis and affecting diagnoses~\cite{granese2023negative}.
Therefore, we aim to verify that the proposed method does not degrade the quality of the reconstructed ECG compared to the original. It is important to note that our primary goal is not to improve ECG quality, as ECGrecover is not intended to be a denoising method.

For this set of analyses, we measured the ECG quality with standard signal quality indicators (SQI): the SQI $_{\text{Avg-QRS}}$~\cite{Makowski2021neurokit} and SQI$_{\text{Fuzzy}}$~\cite{zhao2018sqi} methods, which are both already implemented in the Python library \texttt{Neurokit}~\cite{Makowski2021neurokit}.
SQI$_{\text{Avg-QRS}}$ involves isolating the QRS complexes based on the R-peak locations. Each QRS complex is then standardized by subtracting the average QRS complex and dividing by the standard deviation. The average of these standardized QRS complexes is computed, and the results are re-scaled to a range between 0 and 1, with 1 indicating the highest quality. This method assumes that ECG signal segments exhibit a regular morphology, meaning that most QRS complexes are expected to be similar in shape~\cite{campero2023signal}.
SQI$_{\text{Fuzzy}}$ combines different quality indexes: power spectrum distribution of QRS wave, baseline relative power, and kurtosis of the ECG signal. The overall output is then used to classify the signal into one of three quality levels: \textit{excellent}, \textit{barely acceptable}, or \textit{unacceptable}~\cite{zhao2018sqi}. To numerically quantify the results regarding SQI$_{\text{Fuzzy}}$, we calculated the percentage of ECGs classified as \textit{unacceptable} over the total number of ECGs.
In~\Cref{tab:quality_check}, we present the results of the two metrics computed for both the original ECGs in the test set of the Generepol dataset and those reconstructed with ECGrecover. Regarding SQI$_{\text{Avg-QRS}}$, the reconstructed signals maintain, on average, the same quality as the original signals, with fluctuations of $\sim$0.01 between the results of the two methods. 
In contrast, SQI$_{\text{Fuzzy}}$ shows a lower proportion of ECGs labelled as \textit{unacceptable} with ECGrecover compared to the original dataset. 
This is evident in~\Cref{fig:figure_ecg_example2}, where we observe that ECGrecover attenuates overall noise in the signal more effectively than competitors in the C3 setting. Moreover, in the lead-reconstruction setting, we notice a positive influence from the other leads on the reconstructed one. This effect is illustrated in~\Cref{fig:figure_ecg_example} where the correction of the basal drift is performed for signal in the C$_\text{I}$ setting.

Finally, we would like to emphasize that no additional noise was added to the signals in our comparison, as the proposed method is not intended to be a denoising technique. Additionally, it is important to note that the concept of ECG quality is not perfectly defined and may vary depending on the measurement criteria. Nevertheless, these results support the usability of ECGrecover as a reconstruction method in real-life settings.

\subsection{Random masking}
\label{app:rnd_msk}
\begin{table*}[!htbp]
    \centering
    \caption{Evaluation of ECGRecover and the 
    competitors in terms of PCC and DTW  (mean$\pm$standard deviation) in case of  C$_\text{Rdm}$.}

    \begin{tabular}{lr|ccc|ccc}
    \toprule
      & & \multicolumn{3}{c}{PCC  ($\uparrow$)}   & \multicolumn{3}{|c}{DTW  ($\downarrow$)}\\
\cmidrule(lr){1-2}\cmidrule(lr){3-5}\cmidrule(lr){6-8}
                 \multicolumn{2}{c|}{Lead}  & ECGrecover    & EKGAN~\cite{joo2023twelve}   & Pix2Pix~\cite{isola2017image} & ECGrecover    & EKGAN~\cite{joo2023twelve}   & Pix2Pix~\cite{isola2017image}\\
\cmidrule(lr){1-2}\cmidrule(lr){3-5}\cmidrule(lr){6-8}
\multirow{5}{*}{\begin{sideways} C$_\text{Rdm}$ \end{sideways}} &
           I & \textbf{0.79}±0.11 & 0.23±0.15 & 0.09±0.11 & \textbf{0.05}±0.05 & 0.37±0.24 & 0.19±0.25\\
          & II & \textbf{0.84}±0.11 & 0.26±0.16 & 0.34±0.17 & \textbf{0.03}±0.05  &	0.13±0.06 &	0.10±0.09\\
          & III & \textbf{0.61}±0.22 & -0.02±0.21 & 0.10±0.19 & \textbf{0.08}±0.07 &	0.46±0.5 &	0.36±0.24\\
          & V1 & \textbf{0.82}±0.17 & 0.22±0.13 & 0.18±0.14 & \textbf{0.03}±0.06 &	0.09±0.08 &	0.08±0.07\\ 
          & V6 & \textbf{0.88}±0.09 & 0.24±0.14 & 0.29±0.14 & \textbf{0.02}±0.04 &	0.12±0.06 &	0.10±0.07\\
    \bottomrule
    \end{tabular}
    \label{tab:metric_random_masking}
\end{table*}

\begin{figure}[!htbp]
    \centering
    \includegraphics[width=\columnwidth]{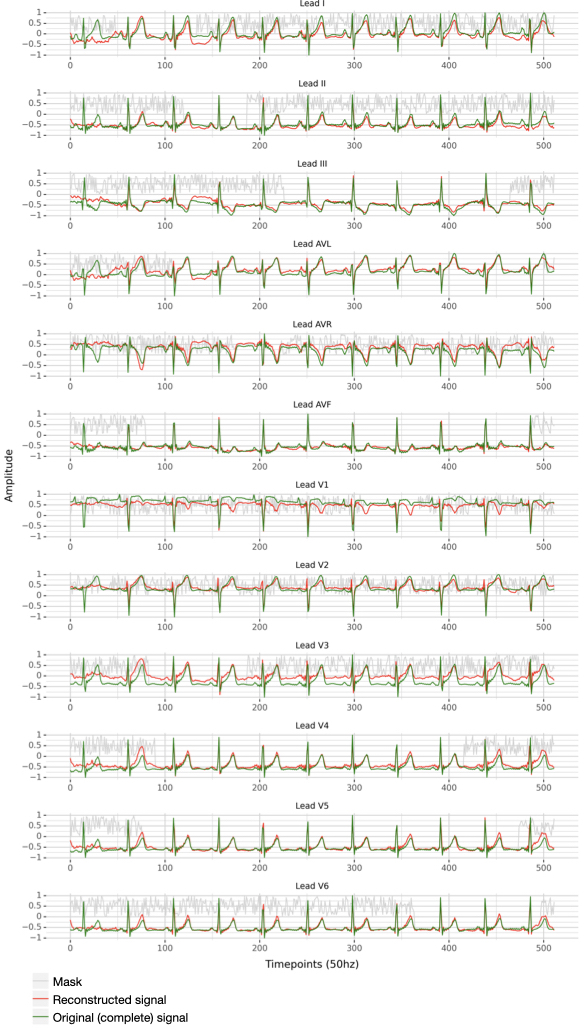}
    \caption{Example of ECG to which we apply a random mask. In green, we indicate the original signals, in red we denote reconstructed ones and in gray we provide the mask. The primer in the figure corresponds to the portion of the gray signals overlapping with the green one.}
    \label{fig:ecg_rd_msk}
\end{figure}
To assess the performance of the proposed ECGrecover, we evaluated its capability to reconstruct ECG signals when subjected to random masking. This entailed applying primers of random duration to randomly chosen parts of each lead, a scenario less likely to occur in real-life settings. This evaluation aimed to verify if the model had learned the inter-lead relationships rather than memorizing specific reconstruction patterns.

An illustration of an ECG subjected to such random masking, C$_\text{Rdm}$, is depicted in \Cref{fig:ecg_rd_msk}. For this process, a random start and end point between 0 and the total length of the signal were selected for each ECG lead. The segments outside the range between these two points were masked.

The results, detailed in \Cref{tab:metric_random_masking}, indicate that ECGrecover consistently outperforms its competitors across various metrics, even in this challenging scenario. For instance, in the most difficult case concerning lead III, ECGrecover not only improves upon the best competitor, Pix2Pix, by 51\% in terms of PCC but also surpasses it by approximately 77.78\% in terms of DTW. Furthermore, ECGrecover demonstrates stable performance across different leads, contrasting with the variability seen in competitors like EKGAN and Pix2Pix, which show wide fluctuations in PCC and DTW scores.

\subsection{Demographic analysis and impact on model performance}
\label{app:Database_info}
\subsubsection{Demographic in Generepol}

\begin{table}[H]
    \caption{Generepol: demographic information summary. We denote by $\mu$ the mean; $\sigma$ the standard deviation; $\eta$ the median. We refer to~\cite{salem2017genome} for further information about the dataset.}
    \footnotesize
    \centering
    \begin{tabular}{r|c|c|c}
    \toprule
     Metric & Age & Height (cm) & Weight (kg) \\
    \midrule
    \multicolumn{4}{l}{\textbf{Training set}: 800 patients (39\% F and 61\% M)} \\
    \cmidrule(lr){1-2}\cmidrule(lr){2-4}
     $\mu\pm\sigma$ & 28$\pm$10 & 170.5$\pm$8.8 & 65.6$\pm$10.1\\
     $\eta$ & 24 & 170 & 65\\
     $\min$ & 18 & 150 & 45\\
     $\max$ & 60 & 197 & 103\\
     \midrule
     \multicolumn{4}{c}{\textbf{Validation set}: 100 patients (40\% F and 60\% M)} \\
    \cmidrule(lr){1-2}\cmidrule(lr){2-4}
      $\mu\pm\sigma$  & 31.5$\pm$12 & 170.1$\pm$9.6 & 66.8$\pm$9.8  \\
     $\eta$ & 25 & 169 &  66\\
     $\min$ & 18 & 152 & 46\\
     $\max$ & 60 & 191 & 95\\
     \midrule
     \multicolumn{4}{c}{\textbf{Testing set}: 89 patients (30\% F and 60\% M)} \\
    \cmidrule(lr){1-2}\cmidrule(lr){2-4}
      $\mu\pm\sigma$  & 25$\pm$8 & 170.4$\pm$7.9 & 64.4$\pm$10\\
     $\eta$ & 22  & 168 & 62\\
     $\min$ & 18 & 151 & 49\\
     $\max$ & 57 & 188 & 102\\
     \bottomrule
    \end{tabular}
    \label{tab:database_info}
\end{table}


We provide in~\Cref{tab:database_info} the summary of demographic information regarding the Generepol dataset. The study included 989 patients with no known heart problems. For a deeper explanation, we refer to~\cite{salem2017genome}.
\subsubsection{Performance of ECGrecover across demographic groups.}
\begin{figure*}[t]
	\centering
		\begin{subfigure}[b]{.6\columnwidth}
		\centering
		\includegraphics[width=\columnwidth]{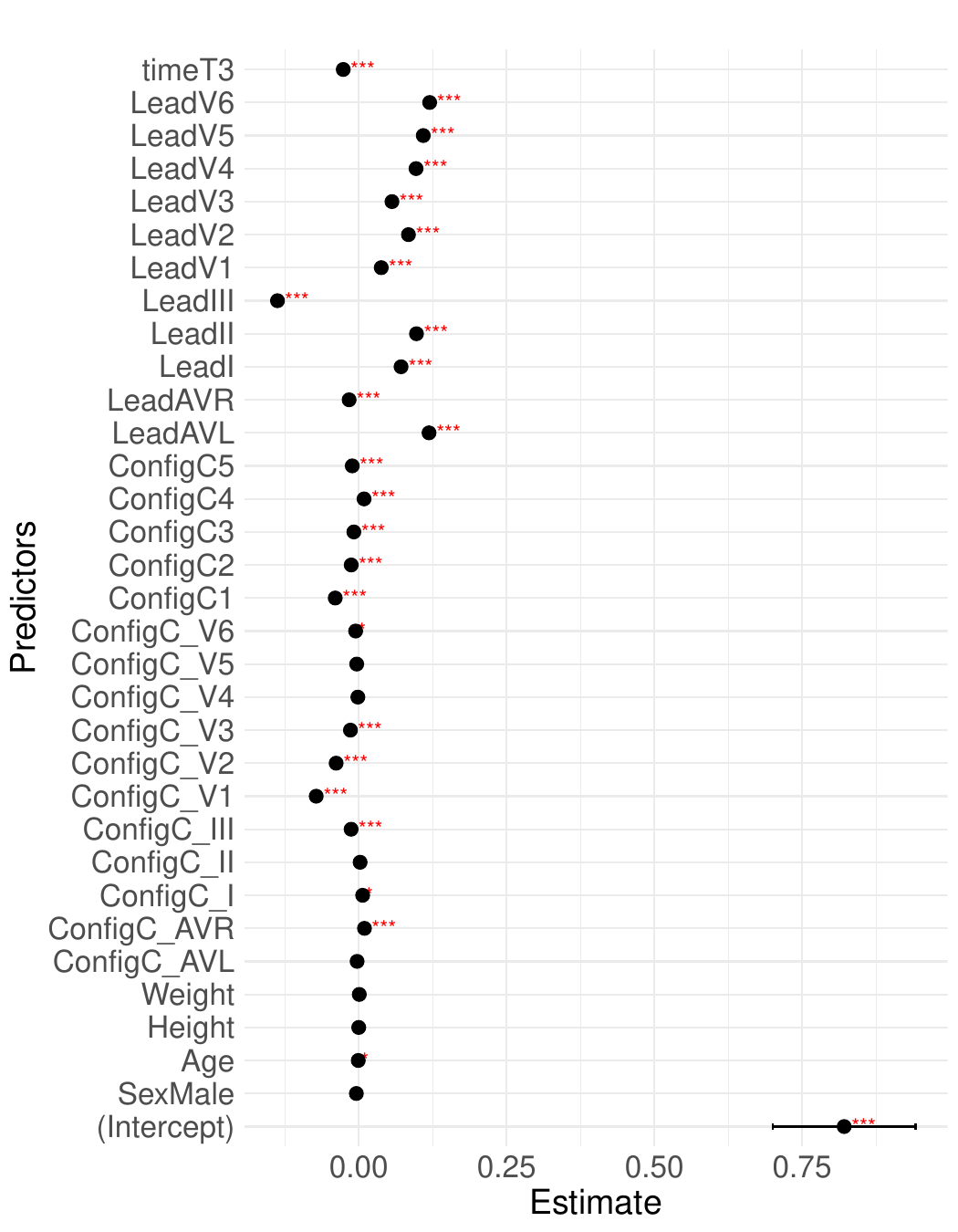}
		\caption{ECGrecover.}
		\label{fig:stats_ecgrecover_pcc}
	\end{subfigure}
 \hfill
 \begin{subfigure}[b]{.6\columnwidth}
		\centering
		\includegraphics[width=\columnwidth]{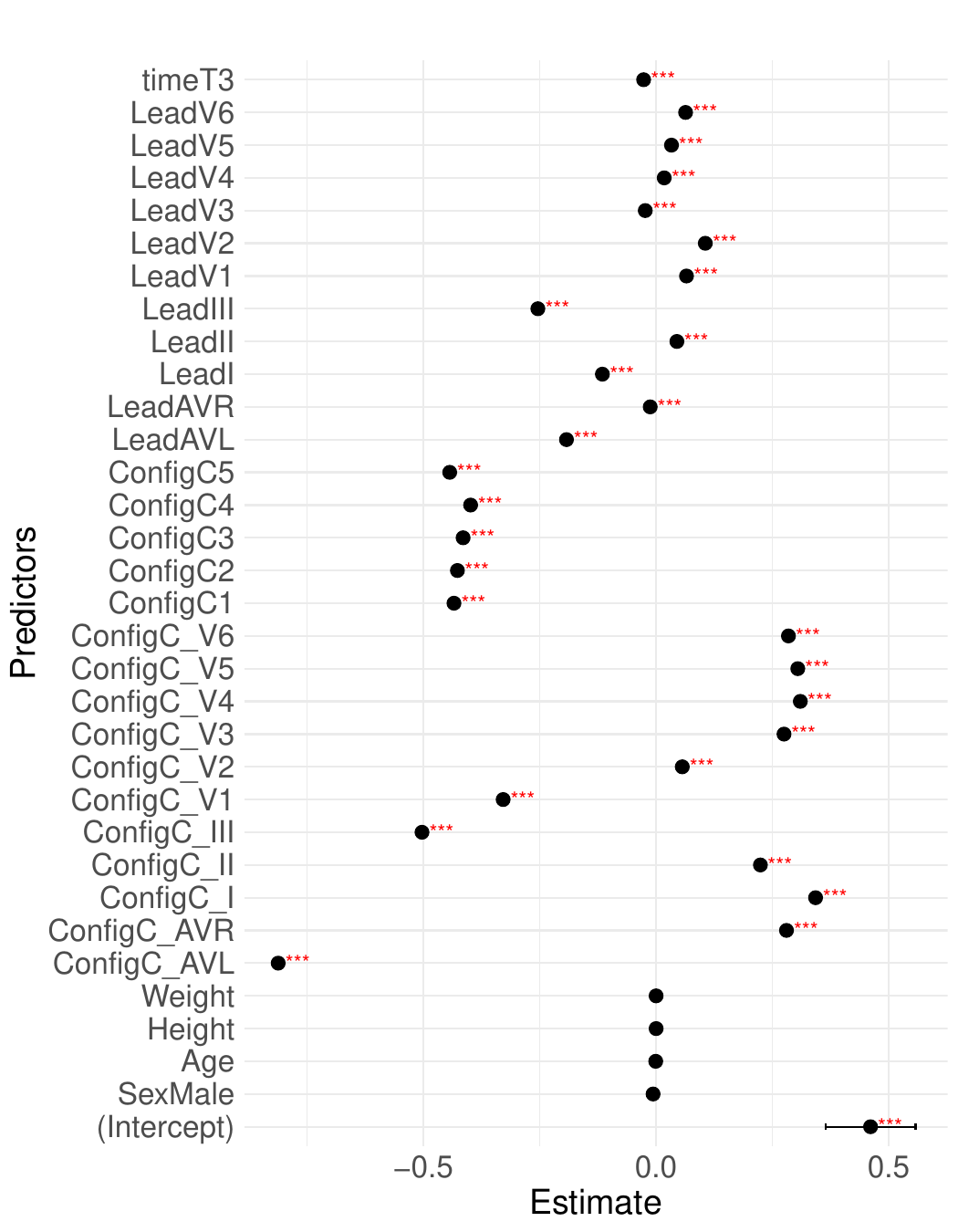}
		\caption{EKGAN.}
		\label{fig:stats_ecgrecover_rmse}
	\end{subfigure}
 \hfill
		\begin{subfigure}[b]{.6\columnwidth}
		\centering
		\includegraphics[width=\columnwidth]{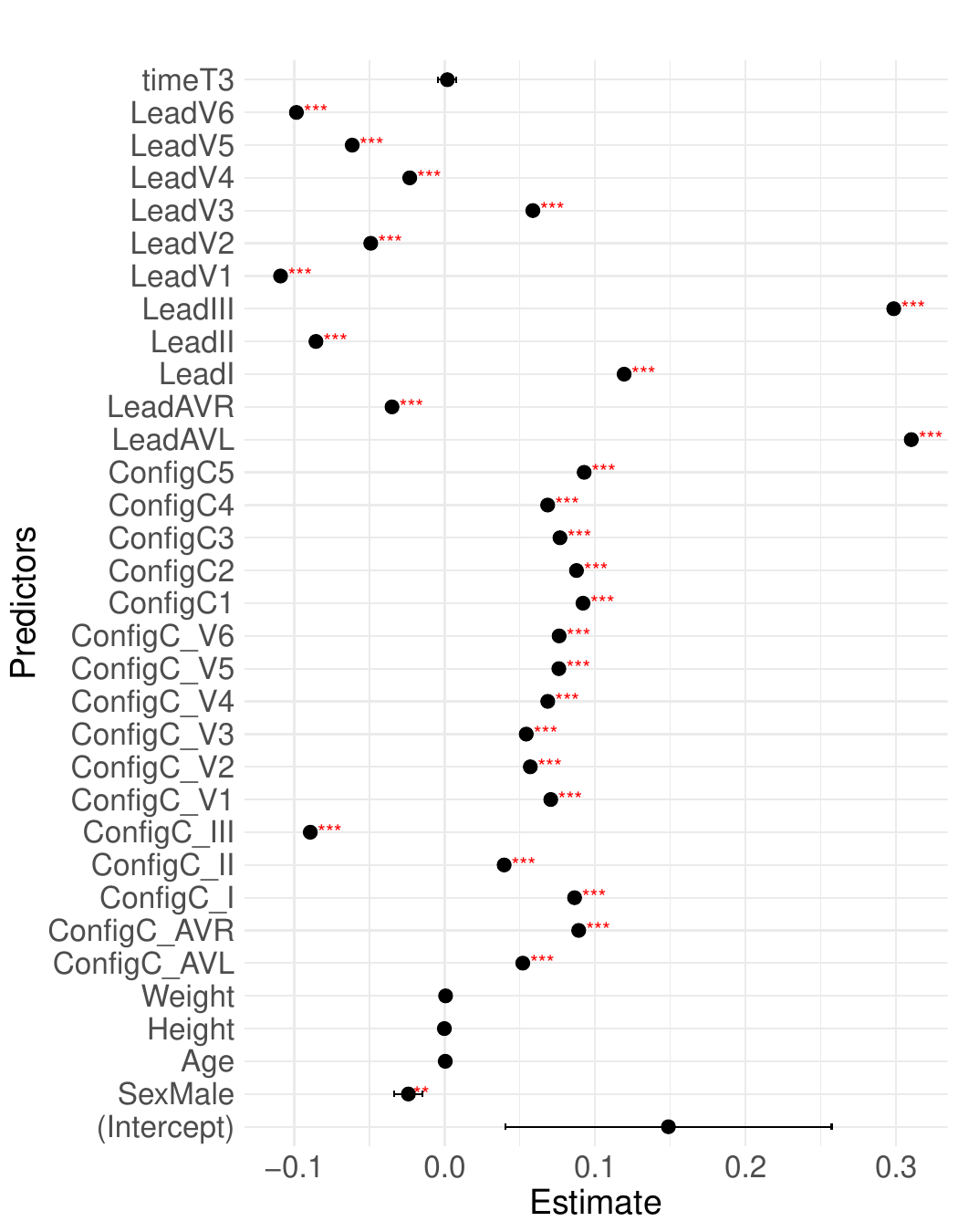}
		\caption{Pix2Pix.}
		\label{fig:stats_pcc}
	\end{subfigure}
	\caption{Coefficients and significance of predictors for PCC. Each row represents a different predictor, and the x-axis indicates the estimate of the coefficient. The red asterisks indicate significance, ($*$) for $p < 0.05$, ($**$) for $p < 0.01$, and ($***$) for $p < 0.001$.}
	\label{fig:stats_pcc}
\end{figure*}
\begin{figure*}[t]
	\centering
		\begin{subfigure}[b]{.6\columnwidth}
		\centering
		\includegraphics[width=\columnwidth]{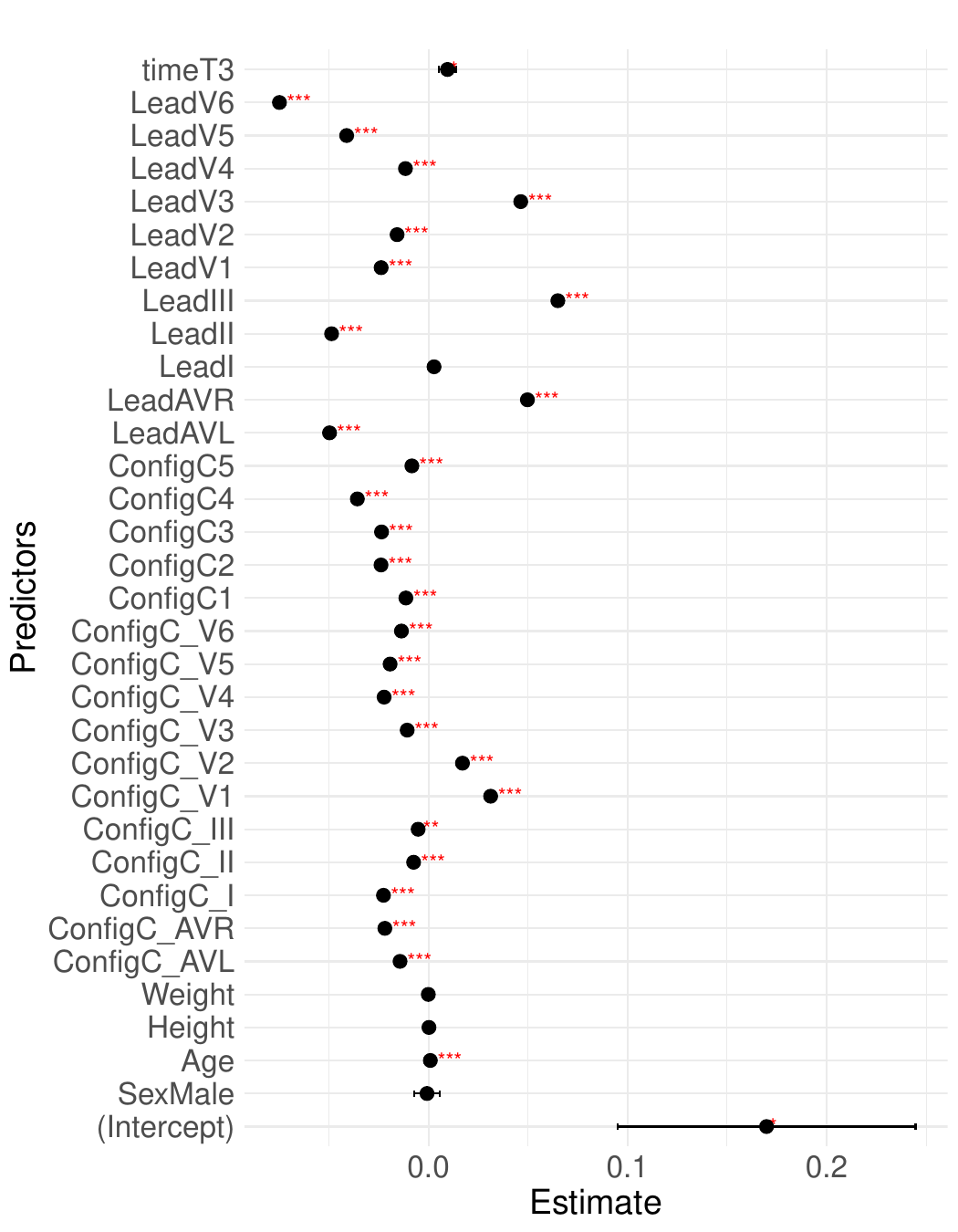}
		\caption{ECGrecover.}
		\label{fig:stats_ecgrecover_rmse}
	\end{subfigure}
 \hfill
 \begin{subfigure}[b]{.6\columnwidth}
		\centering
		\includegraphics[width=\columnwidth]{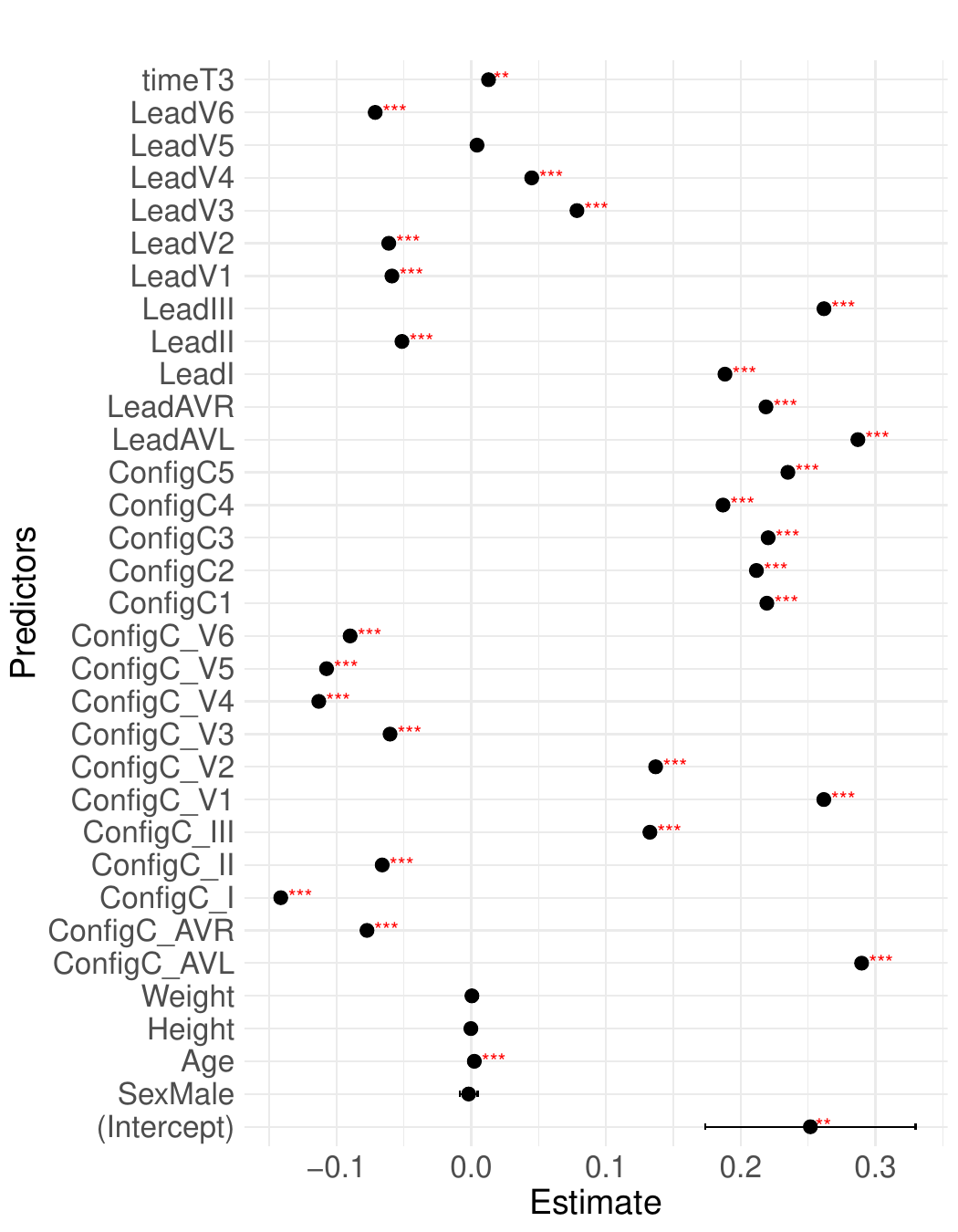}
		\caption{EKGAN.}
		\label{fig:stats_ecgrecover_rmse}
	\end{subfigure}
 \hfill
		\begin{subfigure}[b]{.6\columnwidth}
		\centering
		\includegraphics[width=\columnwidth]{images/Figure8_pix2pix_dtw.pdf}
		\caption{Pix2Pix.}
		\label{fig:stats_rmse}
	\end{subfigure}
	\caption{Coefficients and significance of predictors for RMSE. Each row represents a different predictor, and the x-axis indicates the estimate of the coefficient. The red asterisks indicate significance, ($*$) for $p < 0.05$, ($**$) for $p < 0.01$, and ($***$) for $p < 0.001$.}
	\label{fig:stats_rmse}
\end{figure*}
\begin{figure*}[t]
	\centering
		\begin{subfigure}[b]{.6\columnwidth}
		\centering
		\includegraphics[width=\columnwidth]{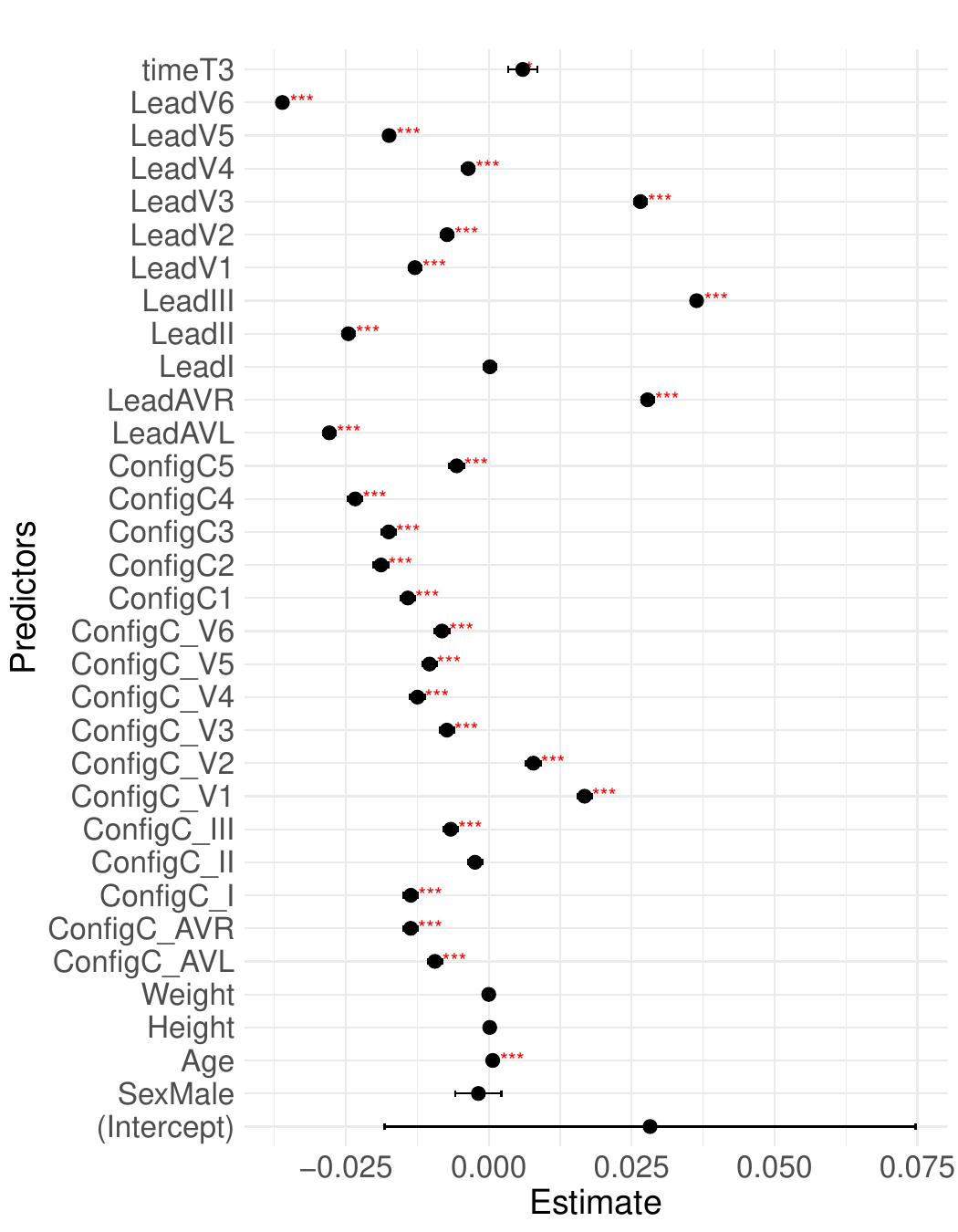}
		\caption{ECGrecover.}
		\label{fig:stats_ecgrecover_dtw}
	\end{subfigure}
 \hfill
 \begin{subfigure}[b]{.6\columnwidth}
		\centering
		\includegraphics[width=\columnwidth]{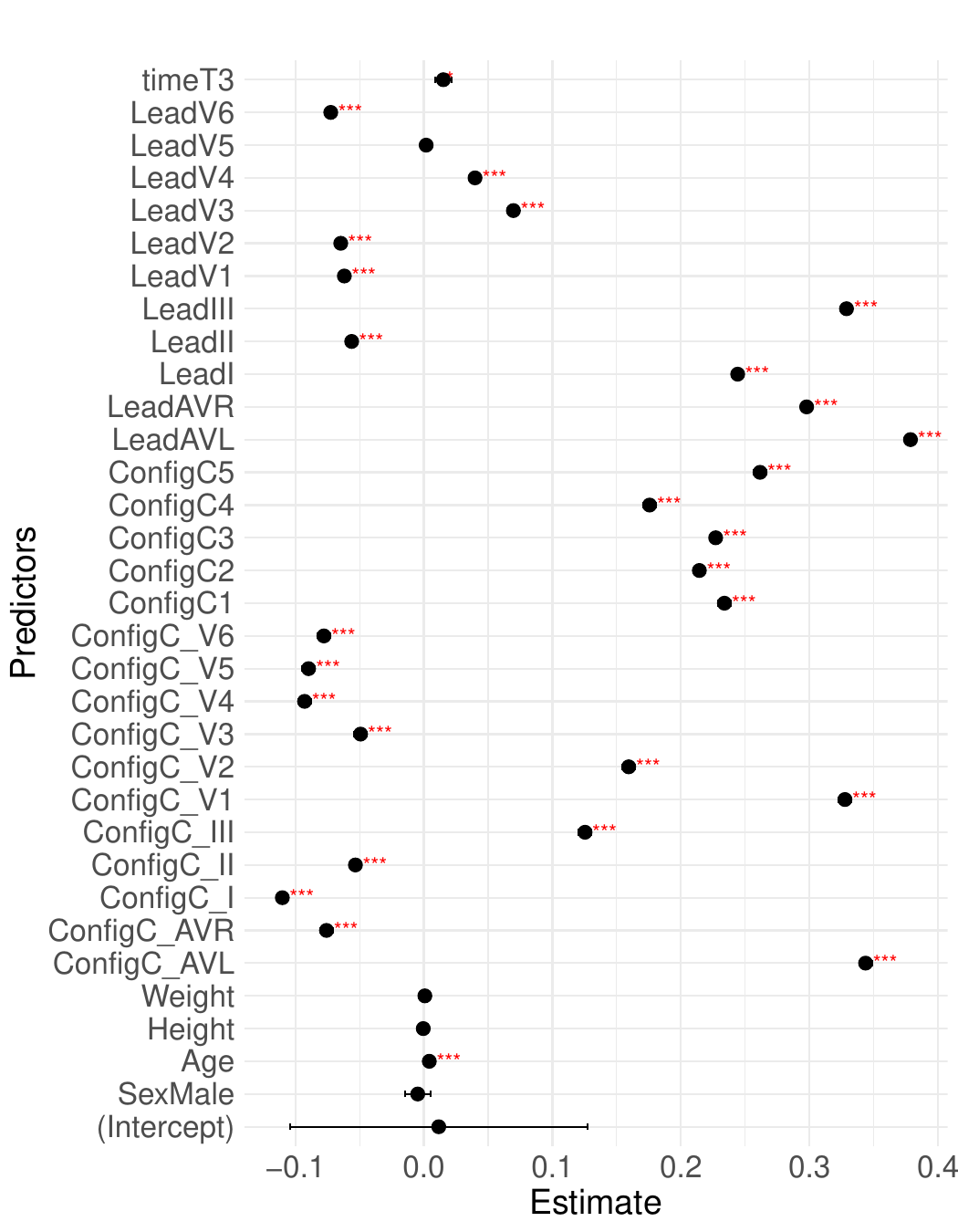}
		\caption{EKGAN.}
		\label{fig:stats_ecgrecover_rmse}
	\end{subfigure}
 \hfill
		\begin{subfigure}[b]{.6\columnwidth}
		\centering
		\includegraphics[width=\columnwidth]{images/Figure8_pix2pix_dtw.pdf}
		\caption{Pix2Pix.}
		\label{fig:stats_dtw}
	\end{subfigure}
	\caption{Coefficients and significance of predictors for DTW. Each row represents a different predictor, and the x-axis indicates the estimate of the coefficient. The red asterisks indicate significance, ($*$) for $p < 0.05$, ($**$) for $p < 0.01$, and ($***$) for $p < 0.001$.}
	\label{fig:stats_dtw}
\end{figure*}
We consider the linear mixed model as our statistical approach to understanding how various predictors (e.g., height, weight, and sex) influence the dependent variables PCC, RMSE, and DTW. In this case, the random effect is the intercept for the ECG. The model was fitted using restricted maximum likelihood and Satterthwaite's method for t-tests.

In \Cref{fig:stats_pcc,fig:stats_rmse,fig:stats_dtw}, we display the statistical analysis results. The x-axis represents the estimated effect size, which measures the relationship between each predictor and the outcome metric. This effect size indicates the expected change in the outcome for a one-unit change in the predictor while keeping all other predictors constant. The distance of each dot from the vertical zero line on the x-axis reflects the magnitude of the effect size, with greater distances signifying stronger effects. Additionally, asterisks next to the dots denote the statistical significance of the estimates: one asterisk ($*$) for $p < 0.05$, two asterisks ($**$) for $p < 0.01$, and three asterisks ($***$) for $p < 0.001$. These asterisks highlight which predictors have a statistically significant impact on the outcome metrics.

The plots show that 'weight' and 'height' do not affect performance metrics across all methods, as their effect size is zero. The same holds for the predictor 'age' concerning the PCC metric, where for ECGrecover, the effect size is very small (<1e-3), cf.~\Cref{fig:stats_pcc}.

Conversely, the estimates for 'age' on RMSE and DTW are very small (close to zero) but statistically significant, indicating a very small but real effect of 'age' on these variables that is not due to random variation (cf.~\Cref{fig:stats_rmse,fig:stats_dtw}). This observation holds for ECGrecover and EKGAN but not for Pix2Pix.

Furthermore, we observed that the time of sotalol administration (3 hours after drug intake) significantly impacts all performance metrics, with a higher effect size (e.g., greater than a 0.01 decrease in PCC). We hypothesize that the drug's footprint or 'age' at a lower level makes it more challenging for all the methods to reconstruct the original ECG from the masked signal.

Finally, the 'sex' predictor shows a slightly negative effect for Pix2Pix across all metrics, indicating poorer performance in male participants.

\subsection{ECGrecover on PTB-XL~\cite{wagner2020ptb}}
\label{app:rst_ptb}
\begin{figure*}[b]
  \centering
  \includegraphics[width=2\columnwidth]{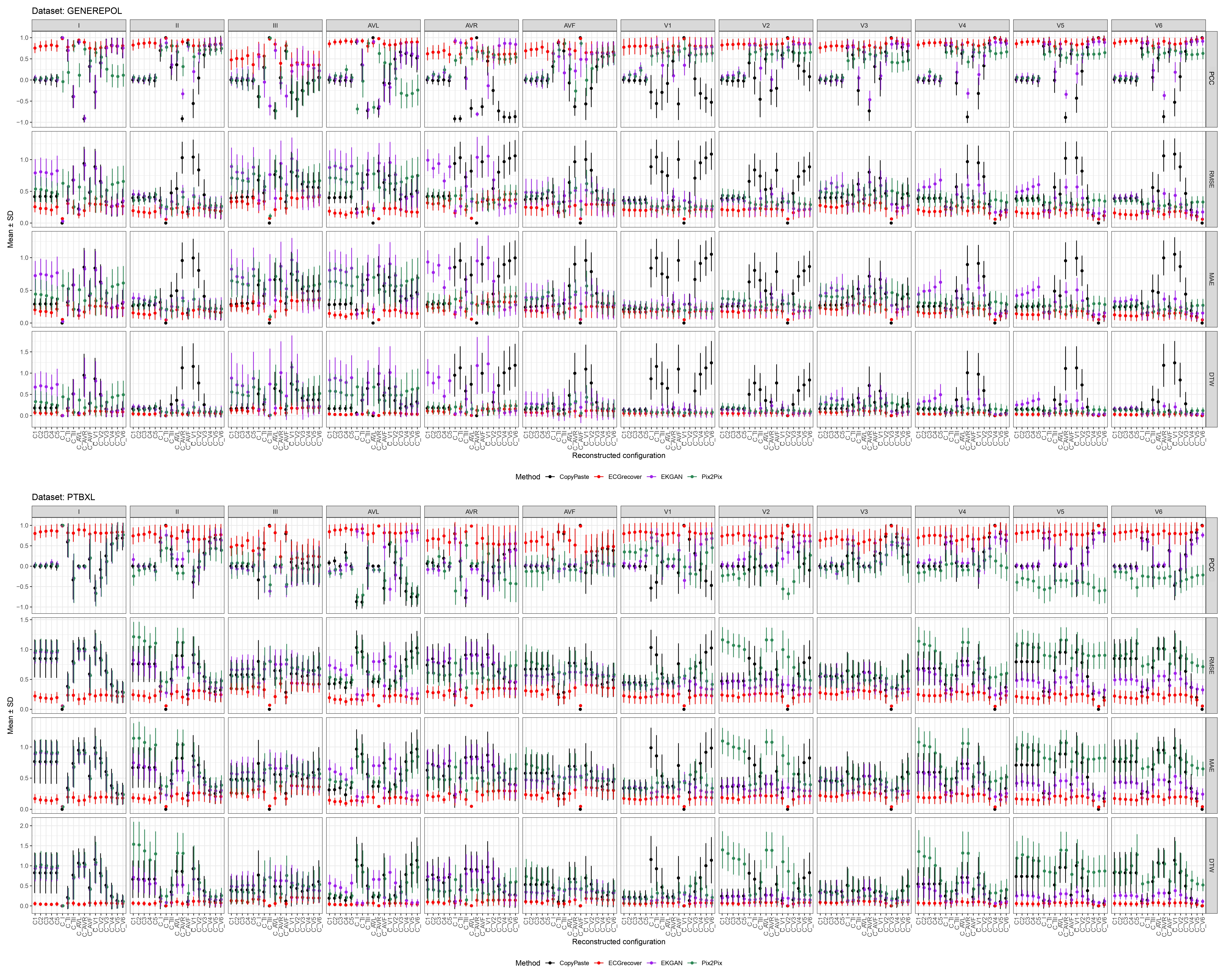}
  \caption{Comprehensive set of results on Generepol~\cite{salem2017genome} and PTB-XL~\cite{wagner2020ptb} datasets.}
  \label{fig:metrics_2_dataset}
\end{figure*}

We present the evaluation of the proposed ECGrecover on PTB-XL~\cite{wagner2020ptb}. Specifically, the dataset contains 21799 recordings from 18869 patients of 10-second 12-
lead ECGs sampled at 500Hz labeled as \textit{normal}, \textit{myocardial} \textit{infarction}, \textit{ST/T change}, \textit{conduction
disturbance} and \textit{hypertrophy}. In~\Cref{fig:metrics_2_dataset}, we show the summary of the results. In particular, we followed the same experimental setting as the one already described for the Generepol dataset in~\Cref{sec:experimental_setup}.To get an overall view, we also include the same representation of the results for the Generepol dataset in the figure.

\subsection{Experimental environment and computational costs}
\label{app:expenv}
In~\Cref{tab:costs}, we summarise the computational time for training and testing ECGrecover and the competitors.
\begin{table}[h]
    \centering
    \caption{The following measures have been run on Tesla P100-PCIE-12GB. In 'Training time', 'Time per ECG' is computed by dividing the time per epoch by the total number of ECGs processed in each epoch.}
    \begin{tabular}{r|c|c|c}
    \toprule
         & ECGrecover & EKGAN~\cite{joo2023twelve} & Pix2Pix~\cite{isola2017image}  \\
         \midrule
         \multicolumn{4}{c}{\textbf{Training time}}\\
         \cmidrule(lr){1-2}\cmidrule(lr){2-4}
         Time per ECG & 4ms & 4ms & 3ms\\
         N. Parameters & 6.147.982 & 26.122.880  &	13.915.393 \\
         GPU memory quota & 0.6 GB &	11.5 GB	& 11.8 GB\\
         \cmidrule(lr){1-2}\cmidrule(lr){2-4}
         \multicolumn{4}{c}{\textbf{Inference time}}\\
         \cmidrule(lr){1-2}\cmidrule(lr){2-4}
         Time per ECG & 2ms & 1ms & 1ms\\
     \bottomrule
    \end{tabular}
    \label{tab:costs}
\end{table}

\section{Comprehensive set of numerical results }
\label{app:res2}
\subsection{Additional plots}
\begin{figure*}[t]
  \centering
  \includegraphics[width=\textwidth]{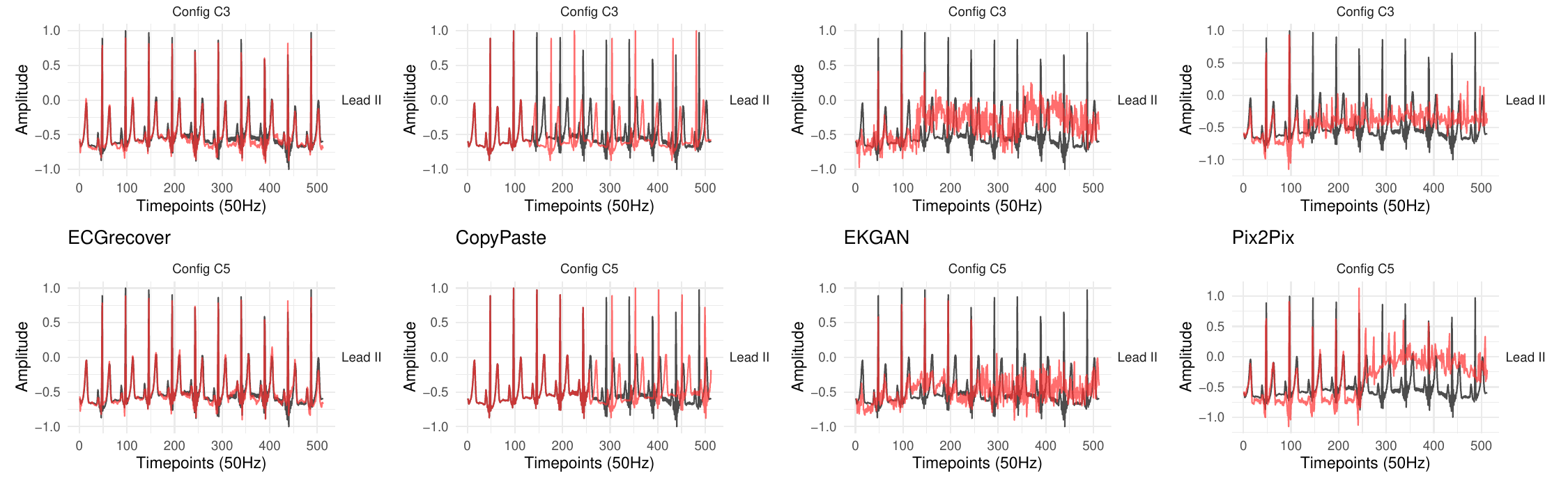}
  \caption{Lead II signals obtained with different approaches for C3 and C5. Reconstructed leads are in red, the original ones are in black. }
  \label{fig:figure_ecg_example2}
\end{figure*}

\begin{figure}[]
    \centering
    \includegraphics[width=.8\columnwidth]{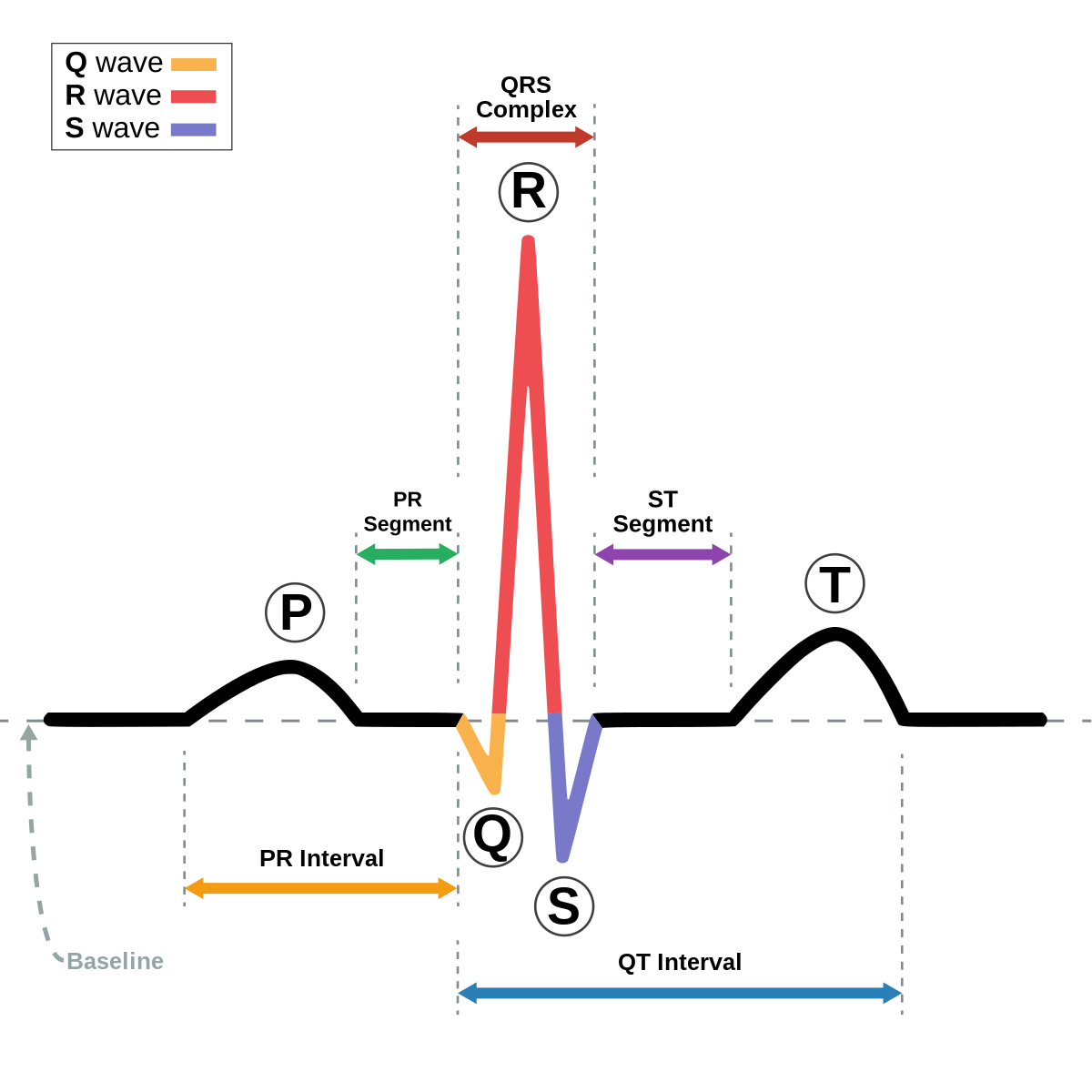}
    \caption{ECG waveforms with areas of interest: P waves, T waves and the QRS complex. From \url{https://fr.wikipedia.org/wiki/Syndrome_du_QT_court}.}
    \label{fig:ecg_peaks}
\end{figure}

We include additional plots in~\Cref{fig:figure_ecg_example2} and~\Cref{fig:ecg_peaks}, which could not be accommodated in the main paper due to space limitations. Specifically, in~\Cref{fig:figure_ecg_example2}, we illustrate a lead II signal reconstruction from input masks C3 and C5 across all methodologies evaluated in our study.   ~\Cref{fig:ecg_peaks} presents the ECG waveforms alongside their labelled wave coordinates (peaks) to enhance comprehension of the findings discussed in~\Cref{sec:res_peaks}.

\subsection{Additional tables}
\label{app:rst_tbls}
This section includes the comprehensive numerical results of the evaluation. Specifically,~\Cref{tab:correlation_sr,tab:correlation_chr,tab:correlation_llr} present the results for the PCC metric;~\Cref{tab:RMSE_sr,tab:RMSE_clr,tab:RMSE_llr} the results for RMSE metric;~\Cref{tab:MAE_sr,tab:MAE_clr,tab:MAE_llr} the results for MAE metric;~\Cref{tab:SDTW_sr,tab:SDTW_clr,tab:SDTW_llr} the results for DTW metric. 

Finally, in~\Cref{tab:QT_sr,tab:QT_chr,tab:QT_llr} we include the results on the QT-segment preservation w.r.t. all the possible input masks. Additionally, we expanded the analysis on the QRS complex (see~\Cref{fig:ecg_peaks}) in~\Cref{tab:QRS_sr,tab:QRS_chr,tab:QRS_llr}. Moreover, in~\Cref{tab:acc_sr,tab:acc_chr,tab:acc_llr}, we show the percentage of R peaks detected (\%-R) in the reconstructed signals.

We highlight that the content of the tables has been analyzed in the paper through various plots and summaries of the results. However, for transparency, we also consider providing the full set of numerical results important.


\begin{table}[!htbp]
    \centering
    \footnotesize
    \caption{ECG segment-recovery: PCC results for all the leads.}
\begin{tabular}{lr|cccc}
    \toprule
      & & \multicolumn{4}{c}{$$PCC$$ ($\uparrow$)} \\
\cmidrule(lr){1-2}\cmidrule(lr){3-6}
                 \multicolumn{2}{c|}{Lead}   & ECGrecover    & EKGAN~\cite{joo2023twelve}   & Pix2Pix~\cite{isola2017image} & CopyPaste \\
    \cmidrule(lr){1-2} \cmidrule(lr){3-6}
\multirow{12}{*}{\begin{sideways} C1\end{sideways}} &
            I & \textbf{0.78}±0.12 & 0.05±0.07 & 0.02±0.05 & -0.01±0.09 \\
          & II & \textbf{0.87}±0.13 & 0.03±0.09 & 0.06±0.07 & -0.02±0.08 \\
          & III & \textbf{0.53}±0.24 & 0.02±0.16 & -0.0±0.1 & -0.01±0.08  \\
          & AVL & \textbf{0.64}±0.16 & 0.04±0.07 & 0.04±0.06 & -0.02±0.1 \\
          & AVR & \textbf{0.87}±0.09 & 0.05±0.1 & 0.01±0.07 & -0.02±0.1 \\
          & AVF & \textbf{0.74}±0.19 & 0.02±0.12 & 0.02±0.08 & -0.02±0.08 \\
\bottomrule
    \end{tabular}
    \label{tab:correlation_sr}
\end{table}
\begin{table}[!htbp]
    \centering
    \footnotesize
    \caption{ECG (limb) lead-reconstruction: PCC results for all the leads.}
    \begin{tabular}{lr|cccc}
    \toprule
      & & \multicolumn{4}{c}{$$PCC$$ ($\uparrow$)} \\
\cmidrule(lr){1-2}\cmidrule(lr){3-6}
                 \multicolumn{2}{c|}{Lead}   & ECGrecover    & EKGAN~\cite{joo2023twelve}   & Pix2Pix~\cite{isola2017image} & CopyPaste \\
    \cmidrule(lr){1-2} \cmidrule(lr){3-6}
\multirow{11}{*}{\begin{sideways} C$_{\text{I}}$\end{sideways}} &
             II & \textbf{0.86}±0.16 & \textbf{0.86}±0.18 & 0.74±0.16 & 0.79±0.26 \\
            & III & \textbf{0.69}±0.3 & 0.66±0.31 & -0.46±0.28 & -0.44±0.27 \\
            & AVL & 0.93±0.07 & \textbf{0.95}±0.08 & 0.47±0.13 & -0.95±0.09 \\
            & AVR & \textbf{0.95}±0.06 & 0.94±0.08 & -0.70±0.12 & 0.91±0.06 \\
            & AVF & 0.70±0.24 & \textbf{0.72}±0.26 & 0.47±0.28 & 0.38±0.38 \\
            & V1 & \textbf{0.87}±0.22 & 0.84±0.22 & 0.63±0.23 & -0.37±0.4 \\
\bottomrule
    \end{tabular}
    \label{tab:correlation_llr}
\end{table}
\begin{table}[!htbp]
    \centering
    \footnotesize
    \caption{ECG (thorachic) lead-reconstruction: PCC results for all the leads.}
    \begin{tabular}{lr|cccc}
    \toprule
      & & \multicolumn{4}{c}{$$PCC$$ ($\uparrow$)} \\
\cmidrule(lr){1-2}\cmidrule(lr){3-6}
                 \multicolumn{2}{c|}{Lead}   & ECGrecover    & EKGAN~\cite{joo2023twelve}   & Pix2Pix~\cite{isola2017image} & CopyPaste \\
    \cmidrule(lr){1-2} \cmidrule(lr){3-6}
\multirow{11}{*}{\begin{sideways} C$_{\text{V1}}$\end{sideways}} &
             I & \textbf{0.79}±0.18 & -0.36±0.4 & 0.75±0.25 & -0.37±0.4 \\
& II & \textbf{0.87}±0.17 & 0.18±0.26 & 0.61±0.25 & -0.65±0.34 \\
& III & \textbf{0.41}±0.31 & 0.21±0.31 & -0.36±0.41 & -0.36±0.42 \\
& AVL & 0.60±0.23 & -0.17±0.31 & \textbf{0.69}±0.15 & 0.50±0.34 \\
& AVR & \textbf{0.87}±0.14 & -0.24±0.42 & 0.51±0.4 & -0.13±0.46 \\
& AVF & \textbf{0.71}±0.24 & 0.52±0.2 & -0.04±0.38 & -0.71±0.38 \\
\bottomrule
    \end{tabular}
    \label{tab:correlation_chr}
\end{table}

\begin{table}[!htbp]
    \centering
    \footnotesize
    \caption{ECG segment-recovery: RMSE results for all the leads.}
    \begin{tabular}{lr|cccc}
    \toprule
      & & \multicolumn{4}{c}{$$RMSE$$ ($\downarrow$)} \\
\cmidrule(lr){1-2}\cmidrule(lr){3-6}
                 \multicolumn{2}{c|}{Lead}   & ECGrecover    & EKGAN~\cite{joo2023twelve}   & Pix2Pix~\cite{isola2017image} & CopyPaste \\
    \cmidrule(lr){1-2} \cmidrule(lr){3-6}
\multirow{12}{*}{\begin{sideways} C1\end{sideways}} &
            I & \textbf{0.24}±0.09 & 0.82±0.22 & 0.47±0.21 & 0.40±0.09 \\
& II & \textbf{0.16}±0.09 & 0.45±0.08 & 0.34±0.09 & 0.39±0.07 \\
& III & \textbf{0.32}±0.11 & 0.83±0.3 & 0.74±0.23 & 0.38±0.1 \\
& AVL & \textbf{0.30}±0.1 & 1.03±0.17 & 0.38±0.14 & 0.41±0.07 \\
& AVR & \textbf{0.17}±0.07 & 0.88±0.27 & 0.69±0.27 & 0.38±0.1 \\
& AVF & \textbf{0.21}±0.13 & 0.39±0.22 & 0.40±0.1 & 0.36±0.09 \\
\bottomrule
    \end{tabular}
    \label{tab:RMSE_sr}
\end{table}
\begin{table}[!htbp]
    \centering
    \footnotesize
    \caption{ECG (limb) lead-reconstruction: RMSE results for all the leads.}
    \begin{tabular}{lr|cccc}
    \toprule
      & & \multicolumn{4}{c}{$$RMSE$$ ($\downarrow$)} \\
\cmidrule(lr){1-2}\cmidrule(lr){3-6}
                 \multicolumn{2}{c|}{Lead}   & ECGrecover    & EKGAN~\cite{joo2023twelve}   & Pix2Pix~\cite{isola2017image} & CopyPaste \\
    \cmidrule(lr){1-2} \cmidrule(lr){3-6}
\multirow{11}{*}{\begin{sideways} C$_\text{I}$\end{sideways}} &
 II & 0.20±0.13 & \textbf{0.19}±0.17 & 0.23±0.14 & 0.29±0.21 \\
& III & \textbf{0.28}±0.15 & 0.33±0.17 & 0.70±0.26 & 0.64±0.23 \\
& AVR & \textbf{0.15}±0.06 & 0.16±0.1 & 0.41±0.14 & 0.93±0.26 \\
& AVL & \textbf{0.13}±0.08 & 0.14±0.07 & 0.76±0.27 & 0.20±0.1 \\
& AVF & 0.25±0.16 & \textbf{0.24}±0.22 & 0.38±0.19 & 0.43±0.23 \\
& V1 & \textbf{0.19}±0.12 & 0.21±0.16 & 0.24±0.11 & 0.91±0.28 \\
\bottomrule
    \end{tabular}
    \label{tab:RMSE_llr}
\end{table}
\begin{table}[!htbp]
    \centering
    \footnotesize
    \caption{ECG (thorachic) lead-reconstruction: RMSE results for all
the leads.}
    \begin{tabular}{lr|cccc}
    \toprule
      & & \multicolumn{4}{c}{$$RMSE$$ ($\downarrow$)} \\
\cmidrule(lr){1-2}\cmidrule(lr){3-6}
                 \multicolumn{2}{c|}{Lead}   & ECGrecover    & EKGAN~\cite{joo2023twelve}   & Pix2Pix~\cite{isola2017image} & CopyPaste \\
    \cmidrule(lr){1-2} \cmidrule(lr){3-6}
\multirow{11}{*}{\begin{sideways} C$_\text{V1}$\end{sideways}} &
I & \textbf{0.26}±0.13 & 0.88±0.27 & 0.30±0.22 & 0.91±0.28 \\
& II & \textbf{0.19}±0.12 & 0.41±0.13 & 0.39±0.15 & 1.08±0.27 \\
& III & \textbf{0.39}±0.13 & 0.96±0.33 & 0.81±0.3 & 0.82±0.31 \\
& AVL & 0.35±0.14 & 1.09±0.32 & \textbf{0.26}±0.13 & 0.32±0.16 \\
& AVR & \textbf{0.20}±0.1 & 0.96±0.32 & 0.57±0.31 & 0.73±0.29 \\
& AVF & \textbf{0.24}±0.17 & 0.45±0.23 & 0.60±0.22 & 1.03±0.3 \\
\bottomrule
    \end{tabular}
    \label{tab:RMSE_clr}
\end{table}

\begin{table}[!htbp]
    \centering
    \footnotesize
    \caption{ECG segment-recovery: MAE results for all
the leads.}
    \begin{tabular}{lr|cccc}
    \toprule
      & & \multicolumn{4}{c}{MAE ($\downarrow$)} \\
\cmidrule(lr){1-2}\cmidrule(lr){3-6}
                 \multicolumn{2}{c|}{Lead}   & ECGrecover    & EKGAN~\cite{joo2023twelve}   & Pix2Pix~\cite{isola2017image} & CopyPaste \\
    \cmidrule(lr){1-2} \cmidrule(lr){3-6}
\multirow{12}{*}{\begin{sideways} C1\end{sideways}} &
            I & \textbf{0.18}±0.08 & 0.76±0.22 & 0.37±0.23 & 0.27±0.09 \\
& II & \textbf{0.13}±0.08 & 0.38±0.08 & 0.26±0.09 & 0.25±0.07 \\
& III & \textbf{0.25}±0.1 & 0.76±0.31 & 0.67±0.25 & 0.27±0.1 \\
& AVL & \textbf{0.23}±0.08 & 0.97±0.18 & 0.26±0.15 & 0.28±0.06 \\
& AVR & \textbf{0.13}±0.06 & 0.81±0.28 & 0.62±0.29 & 0.27±0.09 \\
& AVF & \textbf{0.16}±0.12 & 0.29±0.23 & 0.34±0.11 & 0.22±0.09 \\
\bottomrule
    \end{tabular}
    \label{tab:MAE_sr}
\end{table}
\begin{table}[!htbp]
    \centering
    \footnotesize
    \caption{ECG (limb) lead-reconstruction: MAE results for all
the leads.}
    \begin{tabular}{lr|cccc}
    \toprule
      & & \multicolumn{4}{c}{$$MAE$$ ($\downarrow$)} \\
\cmidrule(lr){1-2}\cmidrule(lr){3-6}
                 \multicolumn{2}{c|}{Lead}   & ECGrecover    & EKGAN~\cite{joo2023twelve}   & Pix2Pix~\cite{isola2017image} & CopyPaste \\
    \cmidrule(lr){1-2} \cmidrule(lr){3-6}
\multirow{11}{*}{\begin{sideways} C$_{\text{I}}$\end{sideways}} &
 II & 0.17±0.13 & \textbf{0.15}±0.17 & 0.17±0.14 & 0.25±0.22 \\
& III & \textbf{0.24}±0.15 & 0.28±0.16 & 0.61±0.29 & 0.54±0.25 \\
& AVL & \textbf{0.12}±0.06 & 0.13±0.1 & 0.33±0.16 & 0.86±0.29 \\
& AVR & \textbf{0.11}±0.08 & \textbf{0.11}±0.07 & 0.69±0.29 & 0.18±0.1 \\
& AVF & 0.21±0.16 & \textbf{0.19}±0.21 & 0.28±0.19 & 0.36±0.23 \\
& V1 & \textbf{0.15}±0.12 & 0.17±0.15 & 0.19±0.11 & 0.87±0.29 \\
\bottomrule
    \end{tabular}
    \label{tab:MAE_llr}
\end{table}
\begin{table}[!htbp]
    \centering
    \footnotesize
    \caption{ECG (thorachic) lead-reconstruction: MAE results for all
the leads.}
    \begin{tabular}{lr|cccc}
    \toprule
      & & \multicolumn{4}{c}{$$MAE$$ ($\downarrow$)} \\
\cmidrule(lr){1-2}\cmidrule(lr){3-6}
                 \multicolumn{2}{c|}{Lead}   & ECGrecover    & EKGAN~\cite{joo2023twelve}   & Pix2Pix~\cite{isola2017image} & CopyPaste \\
    \cmidrule(lr){1-2} \cmidrule(lr){3-6}
\multirow{11}{*}{\begin{sideways} C$_{\text{V1}}$\end{sideways}} &
I & \textbf{0.21}±0.13 & 0.84±0.29 & 0.25±0.23 & 0.87±0.29 \\
& II & \textbf{0.16}±0.12 & 0.33±0.14 & 0.34±0.16 & 1.04±0.29 \\
& III & \textbf{0.33}±0.13 & 0.92±0.34 & 0.74±0.33 & 0.75±0.34 \\
& AVL & 0.29±0.15 & 1.04±0.34 & \textbf{0.21}±0.13 & 0.24±0.16 \\
& AVR & \textbf{0.16}±0.1 & 0.91±0.33 & 0.53±0.32 & 0.69±0.3 \\
& AVF & \textbf{0.20}±0.16 & 0.40±0.24 & 0.54±0.23 & 1.0±0.32 \\
\bottomrule
    \end{tabular}
    \label{tab:MAE_clr}
\end{table}

\begin{table}[!htbp]
    \centering
    \footnotesize
    \caption{ECG segment-recovery: DTW results for all
the leads.}
    \begin{tabular}{lr|cccc}
    \toprule
      & & \multicolumn{4}{c}{$$DTW$$ ($\downarrow$)} \\
\cmidrule(lr){1-2}\cmidrule(lr){3-6}
                 \multicolumn{2}{c|}{Lead}   & ECGrecover    & EKGAN~\cite{joo2023twelve}   & Pix2Pix~\cite{isola2017image} & CopyPaste \\
    \cmidrule(lr){1-2} \cmidrule(lr){3-6}
\multirow{12}{*}{\begin{sideways} C1\end{sideways}} &
            I & \textbf{0.06}±0.06 & 0.67±0.33 & 0.22±0.26 & 0.16±0.09 \\
& II & \textbf{0.03}±0.05 & 0.20±0.07 & 0.11±0.08 & 0.15±0.07 \\
& III & \textbf{0.10}±0.09 & 0.68±0.59 & 0.54±0.32 & 0.14±0.1 \\
& AVL & \textbf{0.09}±0.07 & 1.06±0.32 & 0.14±0.15 & 0.17±0.06 \\
& AVR & \textbf{0.03}±0.04 & 0.77±0.47 & 0.47±0.41 & 0.15±0.09 \\
& AVF & \textbf{0.04}±0.11 & 0.15±0.29 & 0.16±0.1 & 0.13±0.08 \\
\bottomrule
    \end{tabular}
    \label{tab:SDTW_sr}
\end{table}
\begin{table}[!htbp]
    \centering
    \footnotesize
    \caption{ECG (limb) lead-reconstruction: DTW results for all
the leads.}
    \begin{tabular}{lr|cccc}
    \toprule
      & & \multicolumn{4}{c}{$$DTW$$ ($\downarrow$)} \\
\cmidrule(lr){1-2}\cmidrule(lr){3-6}
                 \multicolumn{2}{c|}{Lead}   & ECGrecover    & EKGAN~\cite{joo2023twelve}   & Pix2Pix~\cite{isola2017image} & CopyPaste \\
    \cmidrule(lr){1-2} \cmidrule(lr){3-6}
\multirow{11}{*}{\begin{sideways} C$_{\text{I}}$\end{sideways}} &
 II & \textbf{0.04}±0.09 & \textbf{0.04}±0.14 & 0.06±0.12 & 0.09±0.2 \\
& III & \textbf{0.08}±0.12 & 0.11±0.15 & 0.49±0.45 & 0.41±0.35 \\
& AVR & \textbf{0.02}±0.03 & \textbf{0.02}±0.06 & 0.17±0.14 & 0.87±0.5 \\
& AVL & \textbf{0.02}±0.04 & \textbf{0.02}±0.03 & 0.57±0.44 & 0.04±0.05 \\
& AVF & \textbf{0.06}±0.15 & \textbf{0.06}±0.23 & 0.14±0.24 & 0.18±0.27 \\
& V1 & \textbf{0.04}±0.09 & \textbf{0.04}±0.14 & 0.06±0.09 & 0.82±0.49 \\
\bottomrule
    \end{tabular}
    \label{tab:SDTW_llr}
\end{table}
\begin{table}[!htbp]
    \centering
    \footnotesize
    \caption{ECG (thorachic) lead-reconstruction: DTW results for all
the leads.}
    \begin{tabular}{lr|cccc}
    \toprule
      & & \multicolumn{4}{c}{$$DTW$$ ($\downarrow$)} \\
\cmidrule(lr){1-2}\cmidrule(lr){3-6}
                 \multicolumn{2}{c|}{Lead}   & ECGrecover    & EKGAN~\cite{joo2023twelve}   & Pix2Pix~\cite{isola2017image} & CopyPaste \\
    \cmidrule(lr){1-2} \cmidrule(lr){3-6}
\multirow{11}{*}{\begin{sideways} C$_{\text{V1}}$\end{sideways}} &
             I & \textbf{0.07}±0.1 & 0.78±0.46 & 0.09±0.23 & 0.82±0.49 \\
& II & \textbf{0.04}±0.09 & 0.17±0.13 & 0.15±0.14 & 1.16±0.54 \\
& III & \textbf{0.15}±0.12 & 0.92±0.74 & 0.65±0.49 & 0.67±0.52 \\
& AVL & 0.12±0.13 & 1.19±0.66 & \textbf{0.07}±0.11 & 0.10±0.15 \\
& AVR & \textbf{0.04}±0.07 & 0.92±0.62 & 0.33±0.45 & 0.54±0.46 \\
& AVF & \textbf{0.06}±0.16 & 0.20±0.35 & 0.36±0.28 & 1.07±0.57 \\
\bottomrule
    \end{tabular}
    \label{tab:SDTW_clr}
\end{table}
\clearpage

\begin{table*}[!htbp]
    \centering
    \footnotesize
    \caption{ECG segment-recovery: difference of QT-distance between real and reconstructed lead overall Sot+ and Sot- ECGs.}
    
    \begin{tabular}{lr|cccccccc}
    \toprule
      & & \multicolumn{4}{c}{ $\Delta$-QT (Sot-) $\downarrow$} & \multicolumn{4}{c}{$\Delta$-QT (Sot+) $\downarrow$}\\
      
\cmidrule(lr){1-2}\cmidrule(lr){3-6} \cmidrule(lr){7-10}
                 \multicolumn{2}{c|}{Lead}  & ECGrecover    & EKGAN~\cite{joo2023twelve}   & Pix2Pix~\cite{isola2017image} & CopyPaste  & ECGrecover    & EKGAN~\cite{joo2023twelve}     & Pix2Pix~\cite{isola2017image}  & CopyPaste\\
    \cmidrule(lr){1-2}\cmidrule(lr){3-6} \cmidrule(lr){7-10}
\multirow{12}{*}{\begin{sideways} C1\end{sideways}} &
I & \textbf{0.01}±0.02 & 0.11±0.06 & 0.09±0.04 & 0.05±0.05 & \textbf{0.02}±0.03 & 0.12±0.07 & 0.11±0.05 & 0.07±0.05 \\
& II & \textbf{0.01}±0.01 & 0.12±0.06 & 0.08±0.04 & 0.04±0.04 & \textbf{0.01}±0.03 & 0.15±0.07 & 0.10±0.05 & 0.06±0.07 \\
& III & \textbf{0.07}±0.07 & 0.11±0.09 & 0.11±0.09 & 0.05±0.08 & \textbf{0.06}±0.06 & 0.11±0.09 & 0.13±0.09 & \textbf{0.06}±0.07 \\
& AVL & \textbf{0.03}±0.06 & 0.04±0.06 & \textbf{0.03}±0.04 & 0.03±0.07 & \textbf{0.03}±0.03 & 0.04±0.06 & \textbf{0.03}±0.04 & \textbf{0.03}±0.06 \\
& AVR & \textbf{0.03}±0.03 & 0.11±0.06 & 0.10±0.06 & 0.04±0.05 & \textbf{0.04}±0.05 & 0.11±0.06 & 0.10±0.06 & 0.05±0.06 \\
& AVF & \textbf{0.02}±0.04 & 0.15±0.07 & 0.09±0.05 & 0.05±0.05 & \textbf{0.04}±0.05 & 0.14±0.07 & 0.10±0.06 & 0.06±0.06 \\
\bottomrule
    \end{tabular}
    \label{tab:QT_sr}
\end{table*}

\begin{table*}[!htbp]
    \centering
    \footnotesize
    \caption{ECG (limb) lead-reconstruction: difference of QT-distance between real and reconstructed lead overall Sot+ and Sot- ECGs.}
    \begin{tabular}{lr|cccccccc}
    \toprule
      & & \multicolumn{4}{c}{ $\Delta$-QT (Sot-) $\downarrow$} & \multicolumn{4}{c}{$\Delta$-QT (Sot+)  $\downarrow$}\\
\cmidrule(lr){1-2}\cmidrule(lr){3-6} \cmidrule(lr){7-10}
                 \multicolumn{2}{c|}{Lead}  & ECGrecover    & EKGAN~\cite{joo2023twelve}   & Pix2Pix~\cite{isola2017image} & CopyPaste  & ECGrecover    & EKGAN~\cite{joo2023twelve}     & Pix2Pix~\cite{isola2017image}  & CopyPaste\\
    \cmidrule(lr){1-2}\cmidrule(lr){3-6} \cmidrule(lr){7-10}
\multirow{12}{*}{\begin{sideways} C$_{\text{I}}$\end{sideways}} &
II & \textbf{0.01}±0.01 & \textbf{0.01}±0.01 & \textbf{0.01}±0.03 & \textbf{0.01}±0.02 & \textbf{0.01}±0.03 & \textbf{0.01}±0.03 & 0.02±0.04 & \textbf{0.01}±0.03 \\
& III & \textbf{0.06}±0.07 & 0.08±0.08 & 0.09±0.08 & 0.08±0.07 & \textbf{0.06}±0.07 & 0.07±0.07 & 0.09±0.07 & 0.09±0.07 \\
& AVL & 0.03±0.07 & \textbf{0.02}±0.05 & 0.04±0.04 & 0.14±0.04 & \textbf{0.02}±0.03 & \textbf{0.02}±0.05 & 0.04±0.04 & 0.16±0.05 \\
& AVR & 0.03±0.03 & \textbf{0.01}±0.03 & 0.08±0.05 & \textbf{0.01}±0.02 & 0.04±0.06 & \textbf{0.01}±0.05 & 0.08±0.05 & \textbf{0.01}±0.03 \\
& AVF & \textbf{0.02}±0.04 & 0.04±0.04 & \textbf{0.02}±0.05 & \textbf{0.02}±0.04 & \textbf{0.03}±0.05 & 0.06±0.06 & \textbf{0.03}±0.05 & \textbf{0.03}±0.05 \\
& V1 & \textbf{0.03}±0.04 & 0.04±0.04 & 0.10±0.05 & 0.09±0.06 & \textbf{0.03}±0.04 & \textbf{0.03}±0.04 & 0.12±0.06 & 0.12±0.06 \\
\bottomrule
    \end{tabular}
    \label{tab:QT_llr}
\end{table*}
\begin{table*}[!htbp]
    \centering
    \footnotesize
    \caption{ECG (thorachic) lead-reconstruction: difference of QT-distance between real and reconstructed lead overall Sot+ and Sot- ECGs.}
    \begin{tabular}{lr|cccccccc}
    \toprule
      & & \multicolumn{4}{c}{ $\Delta$-QT(Sot-) $\downarrow$} & \multicolumn{4}{c}{$\Delta$-QT (Sot+)  $\downarrow$)}\\
\cmidrule(lr){1-2}\cmidrule(lr){3-6} \cmidrule(lr){7-10}
                 \multicolumn{2}{c|}{Lead}  & ECGrecover    & EKGAN~\cite{joo2023twelve}   & Pix2Pix~\cite{isola2017image} & CopyPaste  & ECGrecover    & EKGAN~\cite{joo2023twelve}     & Pix2Pix~\cite{isola2017image}  & CopyPaste\\
    \cmidrule(lr){1-2}\cmidrule(lr){3-6} \cmidrule(lr){7-10}
\multirow{12}{*}{\begin{sideways} C$_{\text{V1}}$\end{sideways}} &
I & \textbf{0.02}±0.02 & 0.10±0.05 & \textbf{0.02}±0.02 & 0.09±0.05 & \textbf{0.02}±0.04 & 0.11±0.05 & \textbf{0.02}±0.04 & 0.11±0.05 \\
& II & \textbf{0.01}±0.01 & 0.10±0.05 & 0.02±0.02 & 0.09±0.05 & \textbf{0.02}±0.03 & 0.10±0.04 & 0.03±0.04 & 0.12±0.06 \\
& III & 0.10±0.08 & 0.08±0.06 & \textbf{0.07}±0.07 & \textbf{0.07}±0.07 & 0.10±0.08 & \textbf{0.08}±0.06 & 0.10±0.09 & 0.09±0.08 \\
& AVL & \textbf{0.02}±0.06 & 0.10±0.05 & 0.03±0.03 & 0.04±0.05 & \textbf{0.02}±0.03 & 0.10±0.06 & 0.03±0.03 & 0.04±0.05 \\
& AVR & 0.03±0.03 & 0.10±0.04 & \textbf{0.02}±0.03 & 0.10±0.06 & 0.03±0.04 & 0.10±0.05 & \textbf{0.02}±0.05 & 0.11±0.06 \\
& AVF & \textbf{0.02}±0.04 & 0.07±0.04 & 0.03±0.04 & 0.09±0.06 & \textbf{0.03}±0.05 & 0.09±0.05 & 0.04±0.05 & 0.11±0.07 \\
\bottomrule
    \end{tabular}
    \label{tab:QT_chr}
\end{table*}

\begin{table*}[!htbp]
    \centering
    \footnotesize
    \caption{ECG segment-recovery: difference of QRS-complex between real and reconstructed lead overall Sot+ and Sot- ECGs.}
    \begin{tabular}{lr|cccccccc}
    \toprule
      & & \multicolumn{4}{c}{ $\Delta$-QRS (Sot-) $\downarrow$} & \multicolumn{4}{c}{$\Delta$-QRS (Sot+) $\downarrow$}\\
\cmidrule(lr){1-2}\cmidrule(lr){3-6} \cmidrule(lr){7-10}
                 \multicolumn{2}{c|}{Lead}  & ECGrecover    & EKGAN~\cite{joo2023twelve}   & Pix2Pix~\cite{isola2017image} & CopyPaste  & ECGrecover    & EKGAN~\cite{joo2023twelve}     & Pix2Pix~\cite{isola2017image}  & CopyPaste\\
    \cmidrule(lr){1-2}\cmidrule(lr){3-6} \cmidrule(lr){7-10}
\multirow{12}{*}{\begin{sideways} C1\end{sideways}} &
I & \textbf{0.01}±0.02 & 0.02±0.02 & 0.03±0.02 & 0.03±0.07 & \textbf{0.01}±0.02 & 0.02±0.03 & 0.02±0.02 & 0.03±0.09 \\
& II & \textbf{0.01}±0.01 & 0.02±0.02 & 0.03±0.02 & 0.02±0.06 & \textbf{0.01}±0.02 & 0.02±0.02 & 0.03±0.02 & 0.03±0.1 \\
& III & 0.05±0.04 & 0.04±0.05 & 0.04±0.06 & \textbf{0.03}±0.06 & 0.05±0.05 & \textbf{0.04}±0.06 & 0.05±0.06 & \textbf{0.04}±0.05 \\
& AVL & 0.03±0.03 & 0.04±0.04 & 0.03±0.04 & \textbf{0.02}±0.04 & \textbf{0.03}±0.03 & 0.04±0.05 & \textbf{0.03}±0.05 & \textbf{0.03}±0.04 \\
& AVR & \textbf{0.02}±0.04 & 0.03±0.03 & 0.03±0.03 & \textbf{0.02}±0.05 & \textbf{0.02}±0.04 & \textbf{0.02}±0.03 & 0.03±0.03 & \textbf{0.02}±0.06 \\
& AVF & \textbf{0.02}±0.03 & 0.03±0.04 & 0.03±0.03 & 0.03±0.06 & \textbf{0.02}±0.04 & 0.03±0.04 & \textbf{0.02}±0.04 & 0.03±0.08 \\
\bottomrule
    \end{tabular}
    \label{tab:QRS_sr}
\end{table*}
\begin{table*}[!htbp]
    \centering
    \footnotesize
    \caption{ECG (limb) lead-reconstruction: difference of QRS-complex between real and reconstructed lead overall Sot+ and Sot- ECGs.}
    \begin{tabular}{lr|cccccccc}
    \toprule
      & & \multicolumn{4}{c}{ $\Delta$-QRS (Sot-) $\downarrow$} & \multicolumn{4}{c}{$\Delta$-QRS (Sot+)  $\downarrow$}\\
\cmidrule(lr){1-2}\cmidrule(lr){3-6} \cmidrule(lr){7-10}
                 \multicolumn{2}{c|}{Lead}  & ECGrecover    & EKGAN~\cite{joo2023twelve}   & Pix2Pix~\cite{isola2017image} & CopyPaste  & ECGrecover    & EKGAN~\cite{joo2023twelve}     & Pix2Pix~\cite{isola2017image}  & CopyPaste\\
    \cmidrule(lr){1-2}\cmidrule(lr){3-6} \cmidrule(lr){7-10}
\multirow{12}{*}{\begin{sideways} C$_{\text{I}}$\end{sideways}} &
 II & \textbf{0.01}±0.01 & \textbf{0.01}±0.01 & 0.02±0.03 & \textbf{0.01}±0.03 & \textbf{0.01}±0.02 & \textbf{0.01}±0.02 & 0.02±0.02 & \textbf{0.01}±0.04 \\
& III & \textbf{0.04}±0.05 & \textbf{0.04}±0.04 & 0.05±0.06 & 0.05±0.06 & 0.04±0.06 & \textbf{0.03}±0.04 & 0.05±0.06 & 0.05±0.06 \\
& AVL & 0.03±0.04 & \textbf{0.02}±0.03 & 0.03±0.03 & 0.04±0.05 & 0.03±0.04 & \textbf{0.02}±0.04 & 0.03±0.04 & 0.04±0.05 \\
& AVR & 0.02±0.05 & \textbf{0.01}±0.03 & 0.02±0.03 & \textbf{0.01}±0.03 & 0.02±0.04 & \textbf{0.01}±0.03 & 0.02±0.03 & \textbf{0.01}±0.02 \\
& AVF & \textbf{0.02}±0.03 & \textbf{0.02}±0.03 & \textbf{0.02}±0.04 & \textbf{0.02}±0.04 & \textbf{0.02}±0.05 & \textbf{0.02}±0.04 & \textbf{0.02}±0.05 & 0.03±0.05 \\
& V1 & \textbf{0.01}±0.01 & \textbf{0.01}±0.01 & 0.02±0.02 & 0.02±0.03 & \textbf{0.01}±0.01 & \textbf{0.01}±0.01 & 0.02±0.02 & 0.02±0.04 \\
\bottomrule
    \end{tabular}
    \label{tab:QRS_llr}
\end{table*}
\begin{table*}[!htbp]
    \centering
    \footnotesize
    \caption{ECG (thorachic) lead-reconstruction: difference of QRS-complex between real and reconstructed lead overall Sot+ and Sot- ECGs.}
    \begin{tabular}{lr|cccccccc}
    \toprule
      & & \multicolumn{4}{c}{ $\Delta$-QRS (Sot-) $\downarrow$} & \multicolumn{4}{c}{$\Delta$-QRS (Sot+)  $\downarrow$}\\
\cmidrule(lr){1-2}\cmidrule(lr){3-6} \cmidrule(lr){7-10}
                 \multicolumn{2}{c|}{Lead}  & ECGrecover    & EKGAN~\cite{joo2023twelve}   & Pix2Pix~\cite{isola2017image} & CopyPaste  & ECGrecover    & EKGAN~\cite{joo2023twelve}     & Pix2Pix~\cite{isola2017image}  & CopyPaste\\
    \cmidrule(lr){1-2}\cmidrule(lr){3-6} \cmidrule(lr){7-10}
\multirow{12}{*}{\begin{sideways} C$_{\text{V1}}$\end{sideways}} &
I & \textbf{0.01}±0.02 & 0.02±0.02 & 0.02±0.02 & 0.02±0.02 & \textbf{0.01}±0.02 & 0.02±0.02 & 0.02±0.03 & \textbf{0.01}±0.02 \\
& II & \textbf{0.01}±0.01 & \textbf{0.01}±0.01 &\textbf{ 0.0}1±0.02 & \textbf{0.01}±0.02 & \textbf{0.01}±0.02 & \textbf{0.01}±0.01 & \textbf{0.01}±0.02 & \textbf{0.01}±0.02 \\
& III & 0.06±0.05 & 0.04±0.05 & \textbf{0.03}±0.05 & \textbf{0.03}±0.06 & 0.07±0.05 & \textbf{0.03}±0.05 & \textbf{0.03}±0.05 & 0.04±0.06 \\
& AVL & \textbf{0.03}±0.03 & \textbf{0.03}±0.03 & \textbf{0.03}±0.03 & \textbf{0.03}±0.04 & \textbf{0.03}±0.04 & \textbf{0.03}±0.04 & \textbf{0.03}±0.04 & \textbf{0.03}±0.05 \\
& AVR & \textbf{0.02}±0.04 & \textbf{0.02}±0.03 & 0.04±0.03 & \textbf{0.02}±0.04 & \textbf{0.02}±0.03 & \textbf{0.02}±0.03 & 0.04±0.04 & \textbf{0.02}±0.03 \\
& AVF & \textbf{0.02}±0.03 & \textbf{0.02}±0.03 & \textbf{0.02}±0.03 & \textbf{0.02}±0.03 & \textbf{0.02}±0.04 & \textbf{0.02}±0.04 & \textbf{0.02}±0.04 & 0.02±0.04 \\
\bottomrule
    \end{tabular}
    \label{tab:QRS_chr}
\end{table*}

\begin{table*}[!htbp]
    \centering
    \footnotesize
    \caption{ECG segment-recovery: percentage of detected R peaks in all the reconstructed leads.}
    \begin{tabular}{lr|cccccccc}
    \toprule
      & & \multicolumn{4}{c}{ \%-R (Sot-) $\uparrow$} & \multicolumn{4}{c}{\%-R (Sot+) $\uparrow$} \\
\cmidrule(lr){1-2}\cmidrule(lr){3-6} \cmidrule(lr){7-10}
                 \multicolumn{2}{c|}{Lead}  & ECGrecover    & EKGAN~\cite{joo2023twelve}   & Pix2Pix~\cite{isola2017image} & CopyPaste  & ECGrecover    & EKGAN~\cite{joo2023twelve}     & Pix2Pix~\cite{isola2017image}  & CopyPaste\\
    \cmidrule(lr){1-2}\cmidrule(lr){3-6} \cmidrule(lr){7-10}
\multirow{12}{*}{\begin{sideways} C1\end{sideways}} &
I & \textbf{100}±5.54 & 18.18±11.35 & 41.18±14.65 & 80.0±17.35 & \textbf{100}±9.56 & 16.67±11.48 & 42.86±15.87 & 75.0±17.36 \\
& II & \textbf{100}±3.15 & 16.67±10.89 & 55.56±15.09 & 77.78±15.86 & \textbf{100}±9.5 & 16.67±11.39 & 55.56±15.82 & 77.78±15.16 \\
& III & \textbf{100}±10.35 & 16.67±11.47 & 22.22±14.43 & 75.0±24.25 & \textbf{100}±14.15 & 14.29±11.95 & 20.0±14.55 & 71.43±26.69 \\
& AVL & \textbf{100}±5.92 & 11.11±11.27 & 40.0±16.57 & 70.0±23.64 & \textbf{100}±6.66 & 12.5±12.43 & 37.98±16.78 & 70.0±25.13 \\
& AVR & \textbf{100}±8.16 & 18.18±11.62 & 33.33±14.57 & 80.0±14.66 & \textbf{100}±13.55 & 18.18±11.77 & 33.33±15.15 & 77.78±16.03 \\
& AVF & \textbf{100}±6.52 & 18.18±11.41 & 50.0±14.51 & 77.78±13.77 & \textbf{100}±9.9 & 12.5±12.73 & 50.0±16.38 & 77.78±17.05 \\
\bottomrule
    \end{tabular}
    \label{tab:acc_sr}
\end{table*}
\begin{table*}[!htbp]
    \centering
    \footnotesize
    \caption{ECG (limb) lead-reconstruction: percentage of detected R peaks in all the reconstructed leads.}
    \begin{tabular}{lr|cccccccc}
    \toprule
      & & \multicolumn{4}{c}{ \%-R (Sot-)  $\uparrow$} & \multicolumn{4}{c}{\%-R (Sot+) $\uparrow$} \\
\cmidrule(lr){1-2}\cmidrule(lr){3-6} \cmidrule(lr){7-10}
                 \multicolumn{2}{c|}{Lead}  & ECGrecover    & EKGAN~\cite{joo2023twelve}   & Pix2Pix~\cite{isola2017image} & CopyPaste  & ECGrecover    & EKGAN~\cite{joo2023twelve}     & Pix2Pix~\cite{isola2017image}  & CopyPaste\\
    \cmidrule(lr){1-2}\cmidrule(lr){3-6} \cmidrule(lr){7-10}
\multirow{12}{*}{\begin{sideways} C$_{\text{I}}$\end{sideways}} &
 II & \textbf{100}±3.14 & \textbf{100}±2.38 & \textbf{100}±1.51 & \textbf{100}±2.18 & \textbf{100}±9.64 & \textbf{100}±8.76 & \textbf{100}±7.22 & \textbf{100}±8.21 \\
& III & \textbf{100}±20.91 & \textbf{100}±18.65 & \textbf{100}±5.43 & \textbf{100}±5.21 & \textbf{100}±16.33 & \textbf{100}±15.77 & \textbf{100}±11.15 & \textbf{100}±10.63 \\
& AVL & \textbf{100}±5.56 & \textbf{100}±12.01 & 91.67±12.6 & \textbf{100}±1.17 & \textbf{100}±5.42 & \textbf{100}±9.52 & \textbf{100}±11.11 & \textbf{100}±1.45 \\
& AVR & \textbf{100}±5.12 & \textbf{100}±4.2 & \textbf{100}±16.33 & \textbf{100}±1.88 & \textbf{100}±9.01 & \textbf{100}±7.94 & \textbf{100}±17.58 & \textbf{100}±8.08 \\
& AVF & \textbf{100}±6.47 & \textbf{100}±6.39 & \textbf{100}±1.48 & \textbf{100}±6.01 & \textbf{100}±10.57 & \textbf{100}±10.76 & \textbf{100}±6.84 & \textbf{100}±9.37 \\
& V1 & \textbf{100}±1.78 & \textbf{100}±1.44 & \textbf{100}±6.69 & \textbf{100}±1.17 & \textbf{100}±3.86 & \textbf{100}±1.8 & \textbf{100}±7.47 & \textbf{100}±1.77 \\
\bottomrule
    \end{tabular}
    \label{tab:acc_llr}
\end{table*}
\begin{table*}[!htbp]
    \centering
    \footnotesize
    \caption{ECG (thorachic) lead-reconstruction: percentage of detected R peaks in all the reconstructed leads.}
    \begin{tabular}{lr|cccccccc}
    \toprule
      & & \multicolumn{4}{c}{ \%-R (Sot-) $\uparrow$} & \multicolumn{4}{c}{\%-R (Sot+) $\uparrow$}\\
\cmidrule(lr){1-2}\cmidrule(lr){3-6} \cmidrule(lr){7-10}
                 \multicolumn{2}{c|}{Lead}  & ECGrecover    & EKGAN~\cite{joo2023twelve}   & Pix2Pix~\cite{isola2017image} & CopyPaste  & ECGrecover    & EKGAN~\cite{joo2023twelve}     & Pix2Pix~\cite{isola2017image}  & CopyPaste\\
    \cmidrule(lr){1-2}\cmidrule(lr){3-6} \cmidrule(lr){7-10}
\multirow{12}{*}{\begin{sideways} C$_{\text{V1}}$\end{sideways}} &
I & \textbf{100}±7.81 & \textbf{100}±9.52 & \textbf{100}±8.36 & \textbf{100}±9.62 & \textbf{100}±13.1 & \textbf{100}±14.85 & \textbf{100}±14.52 & \textbf{100}±14.3 \\
& II & \textbf{100}±2.95 & \textbf{100}±2.78 & \textbf{100}±2.85 & \textbf{100}±5.12 & \textbf{100}±11.06 & \textbf{100}±10.55 & \textbf{100}±9.93 & \textbf{100}±10.11 \\
& III & \textbf{100}±10.36 & \textbf{100}±5.25 & \textbf{100}±7.63 & \textbf{100}±6.74 & \textbf{100}±13.57 & \textbf{100}±12.25 & \textbf{100}±10.64 & \textbf{100}±9.76 \\
& AVL & \textbf{100}±4.66 & \textbf{100}±5.72 & 85.71±15.67 & \textbf{100}±4.34 & \textbf{100}±8.0 & \textbf{100}±7.41 & 87.5±13.96 & \textbf{100}±2.81 \\
& AVR & \textbf{100}±12.57 & \textbf{100}±15.68 & \textbf{100}±14.26 & \textbf{100}±14.62 & \textbf{100}±16.71 & \textbf{100}±18.66 & \textbf{100}±16.34 & \textbf{100}±16.47 \\
& AVF & \textbf{100}±6.73 & \textbf{100}±6.24 & \textbf{100}±6.4 & \textbf{100}±5.87 & \textbf{100}±11.16 & \textbf{100}±13.25 & \textbf{100}±12.7 & \textbf{100}±10.51 \\
\bottomrule
    \end{tabular}
    \label{tab:acc_chr}
\end{table*}

%

\end{document}